\documentclass[lettersize,journal]{IEEEtran}

\usepackage{graphicx}  
\usepackage{subcaption}  
\usepackage{enumitem}
\usepackage{listings}
\usepackage{xcolor}
\usepackage{amsthm}
\newtheorem{theorem}{Theorem}

\usepackage{amsmath,amsfonts}
\usepackage{algorithmic}
\usepackage{array}
\usepackage[caption=false,font=footnotesize,labelfont=sf,textfont=sf]{subfig}
\usepackage{textcomp}
\usepackage{stfloats}
\usepackage{url}
\usepackage{verbatim}
\usepackage{graphicx}
\def\BibTeX{{\rm B\kern-.05em{\sc i\kern-.025em b}\kern-.08em
    T\kern-.1667em\lower.7ex\hbox{E}\kern-.125emX}}
\usepackage{balance}
\usepackage{pifont}
\usepackage{makecell}

\renewcommand\textcolor[2]{#2}

\begin{document}

\title{\LARGE Atomic Smart Contract Interoperability with High Efficiency via Cross-Chain Integrated Execution}
\author{
        Chaoyue~Yin,
        Mingzhe~Li,~\IEEEmembership{Member,~IEEE},
        Jin~Zhang,~\IEEEmembership{Member,~IEEE}, 
        You~Lin,
        Qingsong Wei,~\IEEEmembership{Senior Member,~IEEE},
        and Siow Mong Rick Goh,~\IEEEmembership{Senior Member,~IEEE}% <-this % stops a space

\IEEEcompsocitemizethanks{
        \IEEEcompsocthanksitem This work was supported by the National Natural Science Foundation of China (NSFC) under Grant T2495254 and Fang Keng Fellowship.
\IEEEcompsocthanksitem C. Yin is with the Department of Computer Science and Engineering, Southern University of Science and Technology, Shenzhen 518055, China (email: 12432716@mail.sustech.edu.cn).
\IEEEcompsocthanksitem M. Li is with the School of Computing and Information Technology, Great Bay University, Dongguan 523000, China, and with the Institute of High Performance Computing, A*STAR, Singapore (email: mlibn@connect.ust.hk).
\IEEEcompsocthanksitem J. Zhang is with the Department of Computer Science and Engineering, Southern University of Science and Technology, Shenzhen 518055, China, and also
with Peng Cheng Laboratory, Shenzhen 518055, China (email: zhangj4@sustech.edu.cn).
\IEEEcompsocthanksitem Y. Lin is with the Department of Computer Science and Engineering, Southern University of Science and Technology, Shenzhen 518055, China (email: liny2021@mail.sustech.edu.cn).
\IEEEcompsocthanksitem Q. Wei and S. Goh are with the Institute of High Performance Computing (IHPC), Agency for Science, Technology and Research (A*STAR), Singapore (email: wei\_qingsong@ihpc.a-star.edu.sg, gohsm@ihpc.a-star.edu.sg).
% \IEEEcompsocthanksitem J. Zhang is with the Shenzhen Key Laboratory of Safety and Security for Next Generation of Industrial Internet, Department of Computer Science and Engineering, Southern University of Science and Technology, Shenzhen 518055, China (email: zhangj4@sustech.edu.cn).
\IEEEcompsocthanksitem C. Yin and M. Li are the co-first authors.
\IEEEcompsocthanksitem J. Zhang is the corresponding author.
}

}
\markboth{IEEE Transactions on Parallel and Distributed Systems, VOL. XX, NO. XX, XX 2025}%
{Yin \MakeLowercase{\textit{et al.}}: Atomic Smart Contract Interoperability with High Efficiency via Cross-Chain Integrated Execution}



\maketitle

\begin{abstract}
With the development of Ethereum, numerous blockchains compatible with Ethereum's execution environment (i.e., Ethereum Virtual Machine, EVM) have emerged. 
Developers can leverage smart contracts to run various complex decentralized applications on top of blockchains. 
However, the increasing number of EVM-compatible blockchains has introduced significant challenges in cross-chain interoperability, particularly in ensuring efficiency and atomicity for the whole cross-chain application. 
Existing solutions are \emph{either limited in guaranteeing overall atomicity for the cross-chain application, or inefficient due to the need for multiple rounds of cross-chain smart contract execution.}

To address this gap, we propose \texttt{IntegrateX}, an efficient cross-chain interoperability system that ensures the overall atomicity of cross-chain smart contract invocations. 
The core idea is to \emph{deploy the logic required for cross-chain execution onto a single blockchain, where it can be executed in an integrated manner. }
This allows cross-chain applications to perform all cross-chain logic efficiently within the same blockchain. 
\texttt{IntegrateX} consists of a \emph{cross-chain smart contract deployment protocol} and a \emph{cross-chain smart contract integrated execution protocol.}
% two primary protocols: the Hybrid Cross-Chain Smart Contract Deployment Protocol and the Cross-Chain Smart Contract Integrated Execution Protocol. 
The former achieves efficient and secure cross-chain deployment by decoupling smart contract logic from state, and employing an off-chain cross-chain deployment mechanism combined with on-chain cross-chain verification. 
The latter ensures atomicity of cross-chain invocations through a 2PC-based mechanism, and enhances performance through transaction aggregation and fine-grained state lock. 
We implement a prototype of \texttt{IntegrateX}. Extensive experiments demonstrate that it reduces up to 61.2\% latency compared to the state-of-the-art baseline while maintaining low gas consumption.

\end{abstract}


% Copyright 2017 Sergei Tikhomirov, MIT License
% https://github.com/s-tikhomirov/solidity-latex-highlighting/

\definecolor{verylightgray}{rgb}{.97,.97,.97}

\lstdefinelanguage{Solidity}{
	keywords=[1]{anonymous, assembly, assert, balance, break, call, callcode, case, catch, class, constant, continue, constructor, contract, debugger, default, delegatecall, delete, do, else, emit, event, experimental, export, external, false, finally, for, function, gas, if, implements, import, in, indexed, instanceof, interface, internal, is, length, library, log0, log1, log2, log3, log4, memory, modifier, new, payable, pragma, private, protected, public, pure, push, require, return, returns, revert, selfdestruct, send, solidity, storage, struct, suicide, super, switch, then, this, throw, transfer, true, try, typeof, using, value, view, while, with, addmod, ecrecover, keccak256, mulmod, ripemd160, sha256, sha3}, % generic keywords including crypto operations
	keywordstyle=[1]\color{blue}\bfseries,
	keywords=[2]{address, bool, byte, bytes, bytes1, bytes2, bytes3, bytes4, bytes5, bytes6, bytes7, bytes8, bytes9, bytes10, bytes11, bytes12, bytes13, bytes14, bytes15, bytes16, bytes17, bytes18, bytes19, bytes20, bytes21, bytes22, bytes23, bytes24, bytes25, bytes26, bytes27, bytes28, bytes29, bytes30, bytes31, bytes32, enum, int, int8, int16, int24, int32, int40, int48, int56, int64, int72, int80, int88, int96, int104, int112, int120, int128, int136, int144, int152, int160, int168, int176, int184, int192, int200, int208, int216, int224, int232, int240, int248, int256, mapping, string, uint, uint8, uint16, uint24, uint32, uint40, uint48, uint56, uint64, uint72, uint80, uint88, uint96, uint104, uint112, uint120, uint128, uint136, uint144, uint152, uint160, uint168, uint176, uint184, uint192, uint200, uint208, uint216, uint224, uint232, uint240, uint248, uint256, var, void, ether, finney, szabo, wei, days, hours, minutes, seconds, weeks, years},	% types; money and time units
	keywordstyle=[2]\color{teal}\bfseries,
	keywords=[3]{block, blockhash, coinbase, difficulty, gaslimit, number, timestamp, msg, data, gas, sender, sig, value, now, tx, gasprice, origin},	% environment variables
	keywordstyle=[3]\color{violet}\bfseries,
	identifierstyle=\color{black},
	sensitive=true,
	comment=[l]{//},
	morecomment=[s]{/*}{*/},
	commentstyle=\color{gray}\ttfamily,
	stringstyle=\color{red}\ttfamily,
	morestring=[b]',
	morestring=[b]"
}

\lstset{
	language=Solidity,
	backgroundcolor=\color{verylightgray},
	extendedchars=true,
	basicstyle=\footnotesize\ttfamily,
	showstringspaces=false,
	showspaces=false,
	numbers=none,
	numberstyle=\footnotesize,
	numbersep=9pt,
	tabsize=2,
	breaklines=true,
	showtabs=false,
	captionpos=b
}

\section{Introduction} 

\IEEEPARstart{W}{ith} the advent of Bitcoin and Ethereum~\cite{bitcoin,eth}, 
% and the subsequent development of blockchain technology, 
we have witnessed the emergence of an increasing number of programmable blockchains
% capable of running smart contracts
~\cite{belchior2021pastSV,huang2021survey}. 
% Smart contracts are self-executing programs with the terms directly written into code. 
On these programmable blockchains, developers can write and deploy smart contracts to build various complex decentralized applications (dApps, such as DeFi, NFT, etc.~\cite{wenkai2022defiSV,nadini2021mapping}).
Among these programmable blockchains, those that share compatible smart contract execution environment with Ethereum (i.e., Ethereum Virtual Machine, a.k.a. EVM~\cite{ethereum_evm}) dominate the landscape, accounting for over 90\% of the total value locked \cite{chaintvl}. 
With the increasing number of EVM-compatible blockchains and the diversity of dApps running on each chain, the demand for cross-chain dApps has grown significantly~\cite{OU2022OV}. 
Cross-chain dApps refer to dApps that require \emph{\textbf{cross-chain smart contract invocations \emph{(}CCSCI\emph{)}}} and coordinated execution across multiple blockchains \cite{Falazi2024crosschain}.
% However, across numerous EVM-compatible blockchains, it has become an increasingly pressing issue of achieving \emph{atomic and efficient interoperability for \textbf{cross-chain smart contract invocations}}\textbf{ (\emph{CCSCI})}~\cite{Falazi2024crosschain}.

\begin{figure}[t]
    \centering
    \includegraphics[width=0.525\textwidth]{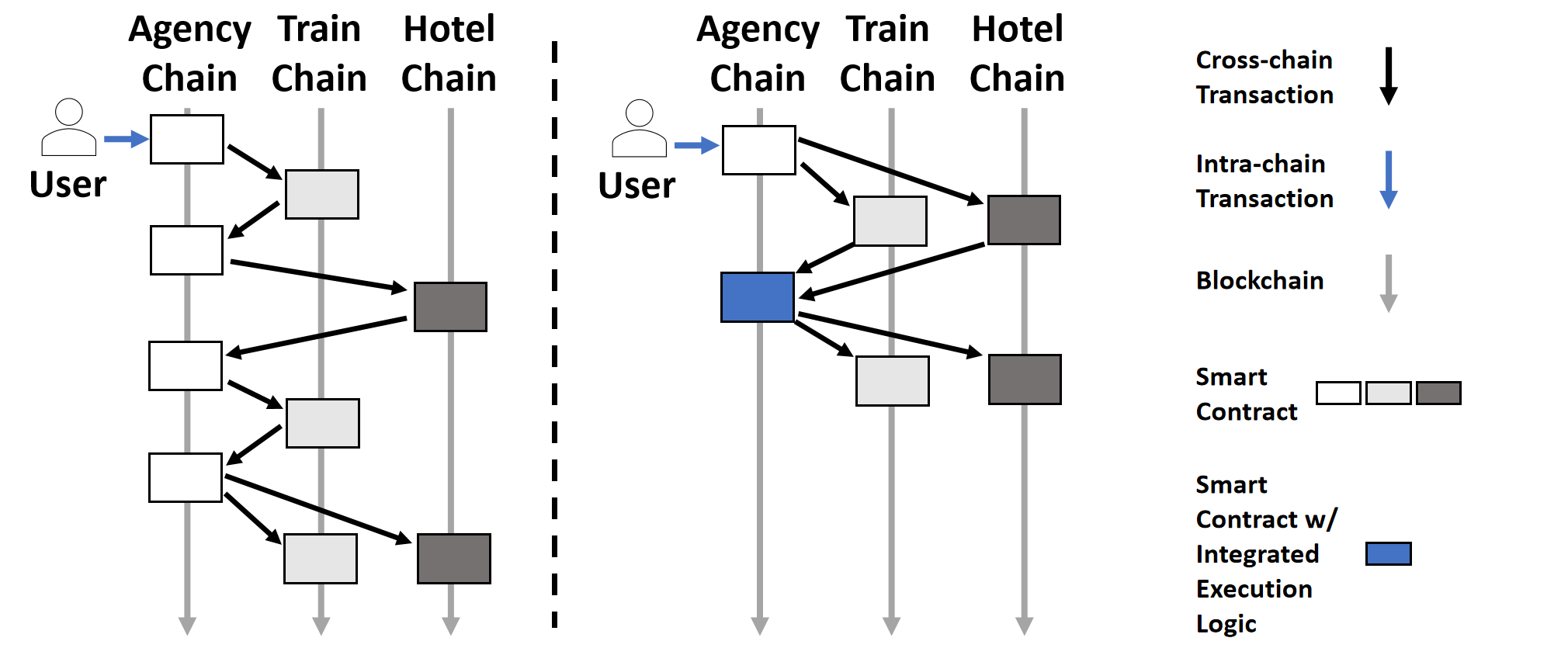}
    \vspace{-18pt}
    \caption{An example of existing CCSCI solutions (left) and \texttt{IntegrateX} (right) in the Train-and-Hotel scenario.}
    % \vspace{-10pt}
    \label{compare}
\end{figure}

However, ensuring overall atomicity for the entire cross-chain dApp while efficiently handling CCSCI remains a critical challenge. 
Consider a classic train-and-hotel problem~\cite{train}, as illustrated in Figure~\ref{compare} left. 
A user wants to book an outbound train ticket (on Train Chain) through a travel agency (on Agency Chain), then book a hotel (on Hotel Chain), followed by a return train ticket (again on Train Chain). 
The user wants to ensure that the entire series of CCSCI either all succeed or all fail, ensuring overall atomicity.
% consider a promising cross-chain flash loan scenario. 
% The user (on Chain A) wants to invoke the flash loan contract (on Chain B) to borrow funds for liquidation on the liquidation contract (on Chain C), and then repay the loan to the flash loan contract (on Chain B).
% Given the strong need for overall atomicity in flash loan scenarios, it is insufficient to only guarantee atomicity for a single cross-chain step. 
% Instead, the entire sequence of CCSCI (borrowing, liquidating, repaying, etc.) must either succeed or fail as a whole (ensuring overall atomicity).
More importantly, maintaining efficiency during CCSCI is crucial. 
% Key considerations include how to minimize latency, reduce gas consumption (monetary cost), and improve concurrency during the CCSCI process.
\textcolor{red}{
Key considerations include minimizing latency, reducing gas consumption (i.e., monetary cost), and improving concurrency during the CCSCI process by avoiding blocking mechanisms such as global locking.}
More cross-chain dApp scenarios are discussed in Section \ref{disscussion}.

Existing CCSCI interoperability solutions generally \emph{either fail to ensure the overall atomicity of a cross-chain dApp or guarantee overall atomicity but with low efficiency. }
To ensure atomicity in the CCSCI process, a widely adopted approach is to use a \emph{two-phase commit (2PC) mechanism \cite{lampson1993twopc, Falazi2024crosschain} involving locking and unlocking}.
For example, some works \cite{nissl2021towards,wood2016polkadot,cosmos2019,abebe2019enabling,darshan2023an,reigsbergen2023demo,ghosh2021leveraging,garoffolo2020zendoo} propose general cross-chain communication protocols that facilitate the transfer of information and data between blockchains through bridging smart contracts deployed on each blockchain.
However, these approaches typically ensure at best the atomicity of single-step cross-chain interactions (Figure \ref{compare}, a single arrow), but fail to guarantee overall atomicity for the entire cross-chain dApp.
Some other recent studies attempt to explore how to ensure overall atomicity for the cross-chain dApp~\cite{robinson2021general,atomic-ibc,chen2024atomci,Falazi2024crosschain}. 
To ensure overall atomicity, the relevant states must remain locked throughout the entire CCSCI process.
However, they commonly face challenges in achieving efficiency. 
% either in maintaining atomicity or in achieving efficiency. 
% For instance, Hyperservice~\cite{liu2021hyperservice} only guarantees financial atomicity, which guarantees the atomicity of the final result by initiating new transactions to re-inject the results. However, it cannot roll back all states and thus cannot ensure the atomicity of a complete cross-chain invocation.
% On the other hand, 
% For instance, works such as GPACT~\cite{robinson2021general} ensure atomicity for entire cross-chain dApps, but their cross-chain interoperability protocols suffer from efficiency issues.
The main reason is that, these approaches usually require \emph{multiple rounds of cross-chain execution and cross-chain information transfer} when handling a cross-chain dApp (Figure \ref{compare}, left), since a cross-chain dApp usually involves interdependent execution logic distributed over multiple blockchains.
% A complex cross-chain dApp typically involves multiple blockchains and several interdependent dApp logic components.
% % , which are implemented through smart contracts. 
% In these works, handling such complexity requires \emph{multiple rounds of cross-chain execution and cross-chain interaction in sequential order across several involved chains. }
% % Specifically, this involves fragmenting the execution logic, reaching consensus, and transferring intermediate states to the next blockchain responsible for executing the subsequent logic fragment. 
It is evident that this method tends to be time-consuming and inefficient (as the locking time could be long), especially when dealing with complex CCSCI. 
More related work (e.g., cross-chain asset swap, smart contract portability) is discussed and differentiated in detail in Section~\ref{related_work}.
% Further related work is discussed in Section~\ref{related_work}.

To fill the research gap, we propose \texttt{IntegrateX}, an \emph{\textbf{efficient} interoperability system that guarantees \textbf{overall atomicity}} for the cross-chain dApp across EVM-compatible blockchains. 
To enhance the efficiency of CCSCI, our core idea is to \emph{clone and deploy\footnote{The logic on the original chain still exists and continues to function normally.} the logic of all contracts involved in a cross-chain dApp—originally distributed across multiple chains—onto a single blockchain}.
% The core idea behind enhancing the efficiency of CCSCI is that, for a cross-chain dApp, \emph{conducting cross-chain deployments for the logic of each involved contract onto a single blockchain.}
This chain thus integrates the entire execution logic of the cross-chain dApp. 
When the CCSCI is required, this chain can perform \emph{\textbf{integrated execution} in one transaction} for all related logic, after receiving the necessary states (Figure \ref{compare}, right). 
% within its environment after receiving the necessary states. 
Since multi-round cross-chain executions and interactions are no longer required, \texttt{IntegrateX} maintains high efficiency, even in complex cross-chain dApps.
% It is important to note that our approach differs from the concept of smart contract portability. 
% In \texttt{IntegrateX}, the states of individual smart contracts are still maintained by their original blockchains. 
% We only “borrow” the execution environment of a single chain to integrate and execute the cross-chain logic. 
% However, the design of \texttt{IntegrateX} is not straightforward, as it faces several significant challenges, which we will address below.

{\textcolor{red}
However, the design of \texttt{IntegrateX} faces several challenges. 

\vspace{3pt}
\noindent
\textbf{Challenge 1:} \emph{How to efficiently and securely deploy smart contracts across chains}. 
Blindly copying full contracts (logic + state) across chains is gas-intensive and vulnerable to byte-code tampering. 
We address this by proposing an on-chain/off-chain \textbf{\emph{Hybrid Cross-Chain Smart Contract Deployment Protocol}}.
Specifically, we \emph{decouple logic from state}, clone only the logic contracts to one target chain (a.k.a. execution chain), and keep state on their original chains. An \emph{off-chain byte-code transfer} cuts gas costs, while an \emph{on-chain hash comparison} between the source and target contracts cryptographically proves correctness.

\vspace{3pt}
\noindent
\textbf{Challenge 2:} \textit{How to achieve high-concurrency, low-overhead during CCSCI while preserving overall atomicity}.
During a cross-chain dApp call, multiple states on different chains must be locked, updated, and released as one atomic unit—without significantly throttling performance. 
We meet this requirement with a 2PC based \textbf{\emph{Cross-Chain Smart Contract Integrated Execution Protocol}}. It (i) locks only the relevant \emph{fine-grained state} slices, (ii) \emph{aggregates multiple cross-chain state transfer} into a single round trip, and (iii) \emph{executes the entire dApp call in one transaction} before unlocking. This design sustains concurrency and keeps latency and gas low even for deep call graphs.
}

This paper mainly makes the following contributions:
% \vspace{-9pt}
\begin{itemize}[left=0pt]
    \item We present \texttt{IntegrateX}, a cross-chain interoperability system that efficiently facilitates CCSCI while ensuring overall atomicity for the cross-chain dApp. 
    \texttt{IntegrateX} can be flexibly deployed on EVM-compatible blockchains without requiring modifications to the underlying blockchain systems. 
    \item In \texttt{IntegrateX}, we propose the Hybrid Cross-Chain Smart Contract Deployment Protocol. 
    It achieves efficient and secure cross-chain deployment through the decoupling of smart contract logic and state, and the hybrid approach of off-chain logic deployment and on-chain comparison verification.
    \item In \texttt{IntegrateX}, we propose the Cross-Chain Smart Contract Integrated Execution Protocol. 
    It ensures overall atomicity of cross-chain invocations through a 2PC-based atomic integrated execution mechanism, and enhances the protocol efficiency through an aggregated cross-chain transaction transmission mechanism and fine-grained state lock.
    \item We implement a prototype of \texttt{IntegrateX} and make it open source~\cite{INTEGRATEX}. 
    Extensive experiments based on real-world use cases demonstrate that \texttt{IntegrateX} reduces latency by up to 61.2\% meanwhile maintains low gas cost and high concurrency compared to the state-of-the-art baseline. 
    In more complex cross-chain invocations, \texttt{IntegrateX} will further improve efficiency.
\end{itemize}
% Contributions. 
% We present \texttt{IntegrateX}, a cross-chain interoperability system that efficiently facilitates cross-chain smart contract invocations while ensuring the atomicity of entire cross-chain dApps. 
% \texttt{IntegrateX} can be flexibly deployed on EVM-compatible blockchains without requiring any modifications to the underlying blockchain systems. 
% Decentralized dApp developers only need to follow our established standards when developing or upgrading their smart contracts to take advantage of \texttt{IntegrateX}'s interoperability features.
% \texttt{IntegrateX} consists of two key protocols. 
% The first is the Hybrid Cross-Chain Smart Contract Deployment Protocol, which achieves efficient and secure cross-chain deployment through the decoupling of smart contract logic and state, and the hybrid approach of off-chain logic deployment and on-chain comparison verification.
% The second is the Cross-Chain Smart Contract Integrated Execution Protocol, which ensures the atomicity of cross-chain invocations through a 2PC-based mechanism, and enhances the protocol efficiency through fine-grained state locks and an aggregated cross-chain transaction transmission mechanism.
% We implement a prototype of \texttt{IntegrateX} and make it open source. 
% Extensive experiments based on real-world use cases demonstrate that \texttt{IntegrateX} delivers exceptional performance.

\textcolor{red}{
The remainder of this paper is organized as follows. Section \ref{related_work} reviews related work on CCSCI solutions. Section \ref{architecture} presents the system model and architecture. Section \ref{HybridP} elaborates on the Hybrid Cross-Chain Smart Contract Deployment Protocol. Section \ref{IntegratedExP} describes the details of the Cross-Chain Smart Contract Integrated Execution Protocol. 
Section \ref{security_analysis} analyzes the security of the proposed system. 
Section \ref{evaluation} evaluates the performance of \texttt{IntegrateX}. Section \ref{disscussion} discusses the limitations and future directions. Finally, Section \ref{conclusion} concludes the paper.
}

\section{Background and Related Work}
\label{related_work}

% \subsection{Cross-Chain Asset Swap and Transfer}

\noindent
\textbf{Cross-Chain Asset Swap and Transfer.}
Blockchain interoperability has gained significant attention. 
% The most extensively researched areas are cross-chain asset swaps and transfers
One widely studied area is cross-chain asset swap and transfer~\cite{2019atomicBEswap,Xu2021htlc,Luo2024crosschannel,Manevich2022ccas,tian2021enabling,herlihy2018atomic,deshpande2020privacy,thyagarajan2022universal,chen2024pacdam,yin2022sidechain,zamyatin2019xclaim}. 
Cross-chain asset swap protocols enable untrusted parties to exchange assets across different blockchain networks, while cross-chain asset transfer protocols involve locking or burning assets on the source chain and creating their equivalents on the target chain. 
% However, these protocols are limited to assets and are not applicable to smart contracts. 
While these protocols provide atomicity guarantees during cross-chain asset swaps or transfers, their protocols cannot be applied to our target scenario of CCSCI.
% Therefore, we propose the \texttt{IntegrateX} protocol, which considers the atomic and efficient cross-chain invocation of smart contracts.

% \subsection{Interoperability Protocol for General Communication}

\vspace{3pt}
\noindent
\textbf{Interoperability Protocol for General Communication.}
Besides cross-chain asset transfers and swaps, some other efforts aim to provide interoperability protocols for more general cross-chain communication~\cite{nissl2021towards,wood2016polkadot,cosmos2019,abebe2019enabling,darshan2023an,reigsbergen2023demo,ghosh2021leveraging,garoffolo2020zendoo}, enabling the transfer of general data beyond assets between blockchains. 
However, the primary limitation of these protocols is that they typically guarantee atomicity for, at most, a single cross-chain step between two blockchains. 
For general cross-chain dApps, which often require multiple rounds of cross-chain interactions across multiple chains, these protocols fail to ensure atomicity for the entire cross-chain dApp. 
In contrast to these approaches, \texttt{IntegrateX} \emph{guarantees overall atomicity of the entire multi-round CCSCI for general cross-chain dApps.}

\vspace{3pt}
\noindent
\textbf{CCSCI with Overall Atomicity.}
% In addition to the aforementioned work, some efforts have been made to address complex cross-chain smart contract invocations (CCSCI) while trying to ensure overall atomicity
There are some studies that attempt to ensure overall atomicity for the cross-chain dApp with complex CCSCI~\cite{liu2021hyperservice,tao2023atomicity,weterkamp2023instant,hu2024ivyredaction,Multi-Chain-Atomic-Commits,robinson2021general,atomic-ibc,chen2024atomci}. 
However, these approaches have various limitations.
For example, HyperService~\cite{liu2021hyperservice} only can provide weak atomicity, named financial atomicity, via an insurance mechanism. 
It fails to provide strong execution atomicity guarantees during CCSCI.
Some works like Heterogeneous Paxos~\cite{Multi-Chain-Atomic-Commits} provide overall atomicity for CCSCI. 
However, they usually require the underlying blockchains to make targeted modifications to fit their protocols, limiting their generality.
Some other works, like Unity~\cite{tao2023atomicity}, also ensure atomicity for a whole CCSCI process.
However, their protocols only apply to cross-chain scenarios between two blockchains and does not take into account the more general multi-chain case.
There are few other solutions that guarantee overall atomicity for the cross-chain dApp with complex CCSCI, \textcolor{red}{as shown in TABLE~\ref{ccsciCop}}. 
However, they generally suffer from inefficiency ~\cite{robinson2021general,atomic-ibc,chen2024atomci}. 
A complex cross-chain dApp typically involves multiple blockchains and several interdependent application logic components (implemented through smart contracts).
In these works, handling such complexity requires multiple rounds of cross-chain execution and cross-chain interaction in sequential order across several involved chains. 
Specifically, this involves locking relevant states, executing the logic fragment, reaching consensus, transferring intermediate states to the next blockchain responsible for executing the subsequent logic fragment, and finally unlocking and updating the states. 
% The main reason is that, these approaches usually require multiple rounds of cross-chain execution and cross-chain information transfer when handling a cross-chain application, since a cross-chain application usually involves interdependent execution logic distributed over multiple blockchains.
As a result, they typically suffer from high latency and poor concurrency handling CCSCI.
{\textcolor{red}For example, Robinson et al. introduced GPACT~\cite{robinson2021general} to achieve atomic cross-chain transactions for Ethereum-based blockchains. However GPACT need to lock entire contracts during updates, reducing availability and performance.
Additionally, during the execution of cross-chain applications, GPACT requires sequentially waiting for each smart contract call to be executed on its respective blockchain and for the results to be transferred across chains, resulting in significant overhead and high latency.}
Additionally, Atomic IBC \cite{atomic-ibc} relies on the Cosmos Hub to provide security guarantees for its protocol. 
This approach is not easily adaptable to other ecosystems (e.g., EVM-compatible blockchains) and may introduce security bottlenecks.
% Robinson et al.~\cite{robinson2021general} introduced GPACT to achieve atomic cross-chain transactions for Ethereum-based blockchains, but it has several limitations.
% Firstly, it locks entire contracts during updates, reducing availability and performance.
% Additionally, during the execution of cross-chain applications, GPACT requires sequentially waiting for each smart contract call to be executed on its respective blockchain and for the results to be transferred across chains, resulting in significant overhead and high latency.
% Atomic IBC~\cite{atomic-ibc} aims to achieve atomicity and low latency in cross-chain applications.
% It provides an efficient and reliable solution for multi-chain interoperability through shared validator sets and atomic transaction bundles.
% By ensuring cross-chain transactions either fully succeed or rollback, and keeping each chain's state unchanged during execution, it optimizes cross-chain communication speed and security, enhancing both developer and user experience.
% However, Atomic IBC faces performance bottlenecks because validators need to join a common engine to atomically execute cross-chain transactions. 
% This centralized execution mechanism can limit the system's scalability and concurrency, potentially becoming a performance bottleneck, especially under high transaction volumes.

Unlike these aforementioned works, \texttt{IntegrateX} is able to maintain high efficiency during CCSCI while ensuring overall atomicity. 
Moreover, it can be deployed to blockchains with the same execution environment (e.g., EVM-compatible) without requiring modifications to the underlying blockchain.

% In the \texttt{IntegrateX} system, we improve the execution efficiency of complex cross-chain smart contract invocations by migrating smart contract logic and utilizing an integrated execution mechanism, thereby avoiding long-term locking of smart contract states. At the same time, we ensure the atomicity of the overall invocation. Most importantly, this protocol does not require modifications to the underlying blockchain protocols or validators to join a shared mempool and blocks, ensuring simplicity of implementation and scalability.

% \subsection{Smart Contract Portability}

\begin{table*}[htbp]
\caption{\textcolor{red}{Comparison of Related Works about CCSCI Execution}}
\label{ccsciCop}
\centering
\textcolor{red}
\begin{tabular}{l|c|c|c|c}
\hline
 & \textbf{Strong Atomicity} & \makecell{\textbf{Compatible with}\\\textbf{Existing Blockchain}}  & \textbf{EVM-compatible}& \textbf{Single-round Execution} \\
\hline
AtomCI\cite{chen2024atomci},Ivyredaction\cite{hu2024ivyredaction},\newline GPACT\cite{robinson2021general},Unity\cite{tao2023atomicity}  & \ding{51} & \ding{51} & \ding{51} & \ding{55} \\
Atomic IBC\cite{atomic-ibc} & \ding{51} & \ding{51} & \ding{55} & \ding{55} \\
Heterogeneous Paxos\cite{Multi-Chain-Atomic-Commits} & \ding{51} & \ding{55} & \ding{55} & \ding{51} \\
HyperService\cite{liu2021hyperservice} & \ding{55} & \ding{55} & \ding{51} & \ding{55} \\
SmartSysc\cite{weterkamp2023instant} & read-only & \ding{51} & \ding{51} & \ding{51} \\
\textbf{\texttt{IntegrateX}} & \ding{51} & \ding{51} & \ding{51} & \ding{51} \\
\hline
\end{tabular}
\label{tab:comparison_double}
\end{table*}

\vspace{3pt}
\noindent
\textbf{Smart Contract Portability.}
Another type of approach, known as smart contract portability~\cite{westerkamp2019verifiable,fynn2020smom,westerkamp2022smartsync}, involves frequently moving or replicating entire smart contracts with all the states associated across different blockchains. 
This allows cross-chain invocations to be converted into intra-chain invocations during contract execution. 
However, the practicality of this approach is often questionable. 
The choice of which blockchain to deploy and run a smart contract and dApp typically involves several considerations, including security, ecosystem, and business concern. 
Moving an entire smart contract from one ecosystem to another is often not favored by developers, as it may compromise security or disrupt the completeness of the ecosystem.
Besides, smart contracts often manage a significant amount of user state, and frequently relocating all of this state across chains incurs substantial overhead. 
This inefficiency further reduces the practicality of these approaches.

The philosophy of \texttt{IntegrateX} is fundamentally different from smart contract portability. 
In \texttt{IntegrateX}, the states involved in each smart contract are still maintained and updated by their original blockchains. 
This allows dApps to continue benefiting from the security and ecosystem of their original blockchain for most of the time. 
During the CCSCI process, we only temporarily lock part of the state relevant to the invocation and transfer it to a single blockchain, where the logic is integrated and executed using that blockchain's execution environment.

% In the \texttt{IntegrateX} system, we migrate smart contract logic so that the logic for complex cross-chain smart contract invocations can be entirely stored on the same chain. Some existing works have also proposed the portability of smart contracts~\cite{westerkamp2019verifiable,fynn2020smom,westerkamp2022smartsync}. However, these works primarily focus on migrating or synchronizing the state of a smart contract from the original chain to another chain, transferring the smart contract itself without involving cross-chain invocations. In \texttt{IntegrateX}, we only migrate the logic of the smart contract, and the migrated contract is used for subsequent complex cross-chain smart contract invocations.
 \section{System Overview}
\label{architecture}

% In this section, we provide an overview of \texttt{IntegrateX}. 
% \texttt{IntegrateX}'s architecture and workflow overview is shown in Figure~\ref{overview} and Figure~\ref{arch}.
% For a detailed analysis of \texttt{IntegrateX}'s security and further discussion, please refer to Section~\ref{security_analysis} and Section~\ref{disscussion}.

\subsection{System Model and Architecture}

{\textcolor{red}
\noindent
\textbf{Infrastructure Components.}
\texttt{IntegrateX} operates over a network of \emph{existing blockchains} interconnected via \emph{off-chain relayers} and \emph{on-chain bridging contracts}. 
Figure~\ref{arch} provides a high-level architecture overview of \texttt{IntegrateX}.
% , showing multiple blockchains each with a deployed bridging contract, and a set of relayer nodes connecting them. 

In \texttt{IntegrateX}, there are $n$ (variable) \emph{\textbf{existing blockchains}} that share the same smart contract execution environment (e.g., EVM-compatible in this paper). 
These blockchains are developed by their respective projects (e.g., Ethereum, BSC, etc.~\cite{eth,bsc_whitepaper}) and operated by their own blockchain nodes, which handle transaction processing and reach consensus within the blockchain. 

\texttt{IntegrateX} also features $m$ (variable) \emph{\textbf{relayers}}.
Relayers are off-chain agents (implemented as independent processes) responsible for listening to events from bridging contracts and relaying information across chains, similar to many mainstream interoperability protocols \cite{atomic-ibc, cosmos2019, sheng2023trustboost, tas2023interchain}. 
Each relayer holds an account (public–private key pair) on each blockchain, and the relayers do not require special privileges; they interact through standard transactions and digitally sign all relayed messages. 
This requires maintaining a sufficient gas balance on each chain, which can be monitored and refilled as needed via external scripts or relayer management services.
To incentivize honest and timely relaying, our system adopts a fee mechanism: users who initiate cross-chain operations are required to pay a cross-chain fee, which is forwarded to the relayer upon successful message delivery. This compensates the relayer for gas costs and operational overhead, while providing an economic incentive aligned with protocol correctness. This model is inspired by, and aligned with, practices in IBC \cite{cosmos2019}, where relayers are independent yet economically motivated.

Additionally, \texttt{IntegrateX}'s \emph{\textbf{bridging smart contracts}} are deployed on each blockchain (serve as on-chain light clients, like in \cite{cosmos2019, atomic-ibc}).
The bridging contracts mainly serve to verify cross-chain transaction proofs, register all smart contracts in \texttt{IntegrateX} that are eligible for cross-chain invocation, and emit events externally to initiate cross-chain actions.
This transport layer of \texttt{IntegrateX} (relayers + bridging contracts) is similar to that of existing interoperability protocols  (e.g., IBC)~\cite{cosmos2019, atomic-ibc}, and provides basic security (ensuring that only valid, agreed-upon cross-chain messages are accepted). 
However, \texttt{IntegrateX} achieves efficiency and overall atomicity at the application layer, which these other protocols do not.

\vspace{3pt}
\noindent
\textbf{Application Layer Components.}
In addition to infrastructure components, we distinguish several roles in the \texttt{IntegrateX} ecosystem's application layer. 
\emph{\textbf{Intra-chain dApp providers}} are \emph{developers} who create traditional single-chain dApps and deploy the corresponding smart contracts on a blockchain of their choice. 
These intra-chain providers are not active participants in cross-chain protocols; rather, they supply reusable contract logic that could later be invoked across chains. 
\emph{\textbf{Cross-Chain dApp providers}} are \emph{developers} who compose multiple intra-chain dApps (potentially on different chains) into a cross-chain dApp. 
A cross-chain dApp typically consists of multiple interrelated smart contracts on different blockchains. 
The cross-chain provider decides which blockchain will serve as the \emph{execution chain} (i.e., the single target chain that will host the integrated execution for their dApp’s cross-chain logic) and which other blockchains will be \emph{invoked chains} supplying state to the execution chain. (In \texttt{IntegrateX}, any EVM-compatible blockchain can be chosen as the execution chain; no custom blockchain is required, as long as the \texttt{IntegrateX} bridging contracts are deployed on it.) 
Finally, \emph{\textbf{end-users}} interact with the deployed cross-chain dApps by sending transactions to one of the blockchains (often the execution chain) to trigger the cross-chain functionality.
}

\subsection{Threat Model}

In \texttt{IntegrateX}, we assume the standard threat model common in cross-chain systems \cite{atomic-ibc, cosmos2019, sheng2023trustboost, tas2023interchain, chen2024atomci, robinson2021general}.
For each blockchain, the proportion of Byzantine nodes is assumed to be less than the fault tolerance threshold $t$ of the respective blockchain network (e.g., for blockchains using BFT-type consensus protocols under partial-synchronous network, $t=1/3$~\cite{pbft}).
% ; for blockchains using Nakamoto-type consensus protocols, $t=1/2$~\cite{ren2019analysis}). 
This ensures the safety and liveness of consensus within each blockchain.
For the relayers, we make only minimal trust assumptions, assuming that at least one relayer is honest and functioning correctly. 
This assumption is consistent with those made in many existing secure interoperability protocols \cite{atomic-ibc, cosmos2019, sheng2023trustboost, tas2023interchain}. 
As for the dApp providers and users—who represent the application layer components—we make no specific threat assumptions, similar to most existing works~\cite{cosmos2019,chen2024atomci,robinson2021general}. 
However, we do discuss common countermeasures for dealing with malicious behavior from these components in Section~\ref{disscussion}.

\subsection{Objective}

\texttt{IntegrateX} aims to achieve the following primary objectives:
\begin{itemize}[left=0pt]
\item \textbf{Efficiency}: During the process of cross-chain smart contract deployment and invocation, \texttt{IntegrateX} seeks to reduce latency, lower gas consumption, and increase transaction concurrency.
\item \textbf{Overall Atomicity}: \texttt{IntegrateX} aims to ensure overall atomicity for cross-chain dApps that require it during CCSCI. This means guaranteeing that the series of CCSCI operations required by cross-chain dApp providers either all succeed or all fail.
\end{itemize}

In addition, \texttt{IntegrateX} aims to possess the following desirable properties. 
\emph{Reliability}: \texttt{IntegrateX} ensures that cross-chain transactions can still be completed even in the presence of malicious relayers. 
\emph{Verifiability}: \texttt{IntegrateX} guarantees that cross-chain transactions can be verified for authenticity, completeness, and validity, even if malicious relayers are involved. 
\emph{Consistency}: \texttt{IntegrateX} ensures that the state changes across the blockchains involved in the cross-chain transaction remain coordinated and consistent. 

The proofs and experiments related to aforementioned properties are detailed in Section~\ref{security_analysis} and Section~\ref{evaluation}.

% \begin{figure}[t]
%     \centering
%     \includegraphics[width=0.48\textwidth]{Figures/relation_1.png}

%     \caption{The architecture of \texttt{IntegrateX}. 
%     % Execution Chain: the chain responsible for integrated execution.
%     % ; invoked chain: the chain where the invoked contracts are deployed.
%     %four roles: intra-chain DApp providers, cross-chain DApp providers, users, and relayer.
%     }
%     %\vspace{-10pt}
%     \label{overview}
% \end{figure} 

\subsection{Overview of the Core Protocols}
{\textcolor{red}
The operation of \texttt{IntegrateX} consists of two main protocols: \emph{Hybrid Cross-Chain Smart Contract Deployment Protocol}, and \emph{Cross-Chain Smart Contract Integrated Execution Protocol}, as illustrated in Figure \ref{arch}.

\begin{figure}[t]
    \centering
    \includegraphics[width=0.48\textwidth]{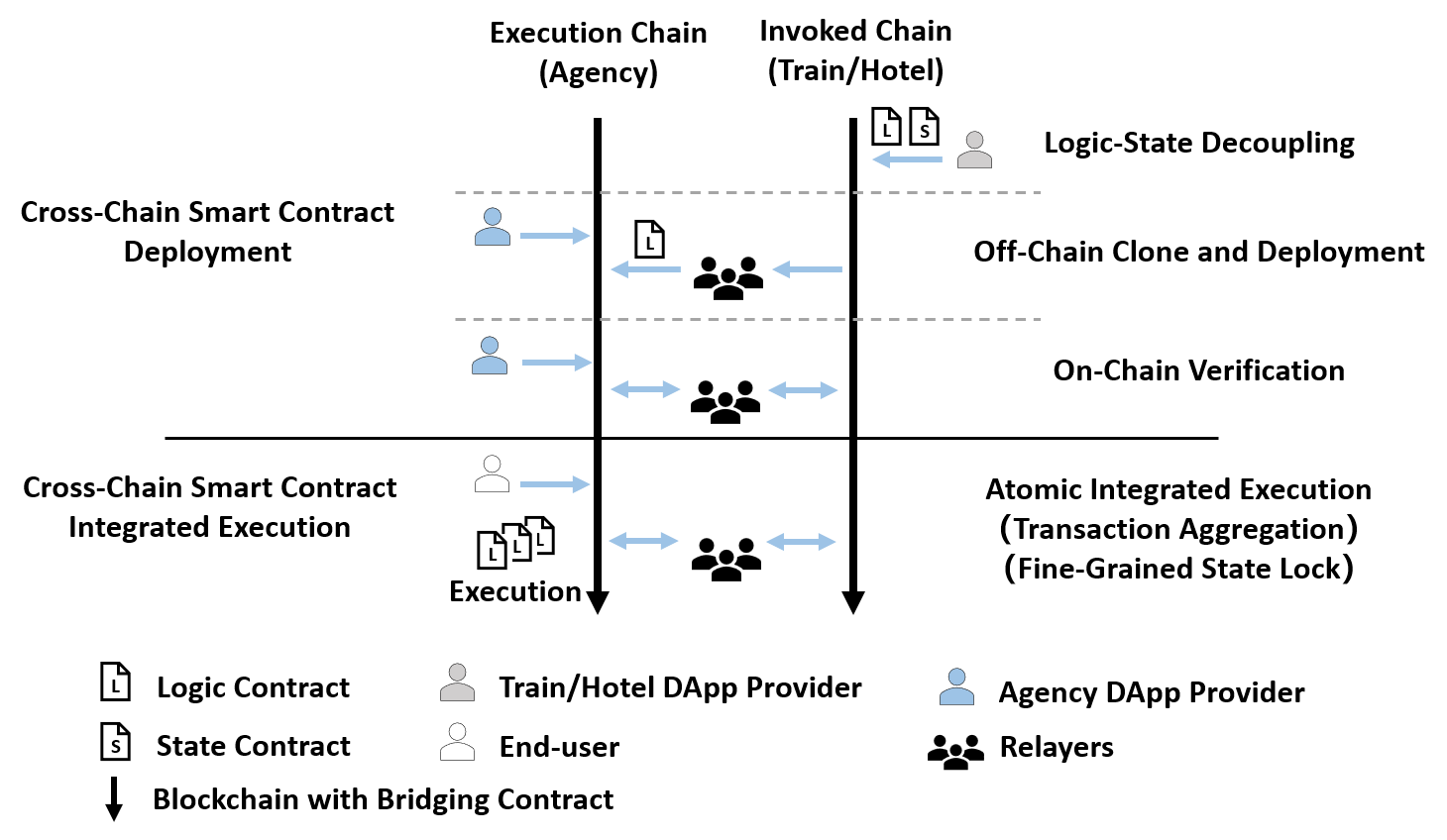}
    \vspace{-9pt}
    \caption{The overview of \texttt{IntegrateX}'s core protocols. 
    % Execution Chain: the chain responsible for integrated execution.
    % ; invoked chain: the chain where the invoked contracts are deployed.
    %four roles: intra-chain DApp providers, cross-chain DApp providers, users, and relayer.
    }
    \label{arch}
\end{figure} 
% The smart contract preparation phase only occurs when dApp providers need to develop or upgrade smart contracts, so the frequency of this phase is relatively low. 
% Similarly, the cross-chain logic contract deployment phase only runs when cross-chain dApp providers need to deploy logic contracts across chains onto the same blockchain, which also occurs infrequently. 
% In contrast, the cross-chain smart contract invocation and integrated execution phase runs every time a user needs to interact with a cross-chain dApp, making it the most frequent phase of operation.

\vspace{3pt}
\noindent
\textbf{Hybrid Cross-Chain Smart Contract Deployment Protocol.} 
To address the high overhead and complexity associated with cross-chain deployment of smart contracts, we design a hybrid deployment protocol that improves efficiency while ensuring security and flexibility. The proposed protocol consists of three core phases: \emph{Logic-State Decoupling}, \emph{Off-Chain Clone and Deployment}, and \emph{On-Chain Verification}, as shown in Figure~\ref{arch}.

The first phase, \emph{Logic-State Decoupling}, tackles the challenge of redundant data migration in cross-chain scenarios. We devise a set of guidelines that enables dApp developers to separate smart contracts into \emph{logic execution contracts} and \emph{state storage contracts}. 
This approach enables our protocol to clone and deploy only the logic contracts onto the execution chain, while the state-heavy contracts remain in their original locations. 
In this phase, dApp providers can flexibly develop logic contracts and state contracts according to our defined logic-state decoupling guidelines (Section \ref{subsec:LSD}). 

To further minimize the gas cost and latency involved in cross-chain logic contract deployment, we design an \emph{Off-Chain Clone and Deployment} mechanism. Instead of performing all operations on-chain, which is resource-intensive, our protocol enables the off-chain cloning and deployment of logic contracts onto the execution chain. This phase is triggered only when dApp providers need to deploy logic contracts from one invoked chain onto the execution chain for execution purposes. The execution chain can be chosen flexibly, and a designated transaction can initiate the deployment process. \texttt{IntegrateX} implements this off-chain deployment technique (Section~\ref{subsec:migration}) to optimize cost-efficiency.

However, off-chain operations pose integrity risks due to their susceptibility to tampering. To ensure correctness and trustworthiness, we introduce an \emph{On-Chain Verification} phase. In this phase, bridging smart contracts, which are deployed on both the invoked chain and the execution chain, are used to perform on-chain cross-verification of the logic contract bytecodes between the two chains. This verification mechanism (Section~\ref{subsec:verification}) guarantees that the deployed contracts are authentic and consistent with their original versions, thereby securing the deployment process against malicious interference.

Through this three-phase protocol design, we effectively reduce deployment overhead while maintaining the integrity and scalability of \texttt{IntegrateX}.

\vspace{3pt}
\noindent
\textbf{Cross-Chain Smart Contract Integrated Execution Protocol.}
To achieve high-concurrency, low-overhead and guarantee atomicity during CCSCI, we design an integrated execution protocol based on the two-phase commit (2PC) paradigm, following mainstream practices (Section~\ref{subsec:execution}).
The protocol coordinates execution by first locking the relevant contract states across participating invoked chains, transferring these states to a designated execution chain, executing the combined logic, and finally returning the updated states to invoked chains for unlocking and commitment. 
This ensures that cross-chain transactions either complete entirely or are rolled back in their entirety, preserving consistency.

However, frequent cross-chain state transfers can introduce substantial communication overhead. To address this challenge, we introduce a transaction aggregation mechanism that consolidates multiple cross-chain interactions into a single transaction when applicable, thereby reducing cross-chain communication costs (Section~\ref{subsec:aggregation}).

Another key issue is the limited concurrency introduced by state locking. In conventional designs, locking entire contract states can hinder parallel transaction processing.
To improve concurrency, our protocol incorporates a fine-grained locking scheme (Section \ref{subsec:lock}). Specifically, we define a set of development guidelines that enable dApp developers to decompose contract states and apply locks at a finer granularity.
By isolating and locking only the necessary sub-states, the protocol enables concurrent execution of unrelated operations, significantly enhancing scalability and throughput in multi-chain environments.

Overall, the protocol ensures atomicity, reduces communication overhead, and improves concurrency, thereby enabling reliable and efficient execution of cross-chain dApps.
}
% This phase runs every time a user needs to interact with a cross-chain dApp, making it the most frequent phase of operation.
% In this phase, users interact with cross-chain dApps by sending transactions to the target chain, invoking cross-chain smart contracts based on the dApp’s application logic. 
% \texttt{IntegrateX} employs an atomic integrated execution scheme (Section \ref{subsec:execution}), similar to the 2PC protocol, to ensure the overall atomicity of the series of CCSCI involved in a cross-chain dApp.
% This scheme first locks the relevant states of the smart contracts involved in the cross-chain dApp on the respective chains at a fine-grained level, 
% then transmits these states across chains to the target chain. 
% Since the target chain already contains all the execution logic required for the cross-chain dApp (from the earlier phase), it can integrate and execute all logic within one transaction once the necessary states have been received. 
% After execution, the new states are returned to the corresponding chains for unlocking and updating.
%Additionally, \texttt{IntegrateX} employs a transaction aggregation mechanism (Section \ref{subsec:aggregation}) during this phase to reduce cross-chain overhead.

% In the subsequent descriptions within this paper, we will refer to the target chain selected by a cross-chain dApp for deployment—i.e., the chain responsible for integrated execution—as the \textbf{\emph{execution chain}}. 
% The other chains associated with the cross-chain dApp will be collectively referred to as \textbf{\emph{invoked chains}}.

\section{Hybrid Cross-Chain Smart Contract Deployment Protocol}
\label{HybridP}
% We propose an on-/off-chain Hybrid Cross-Chain Smart Contract Deployment Protocol to efficiently and securely deploy smart contracts across chains. In this section, we will provide a detailed explanation of the design of the protocol.
\begin{figure}[t]
    \centering
    \includegraphics[width=0.48\textwidth]{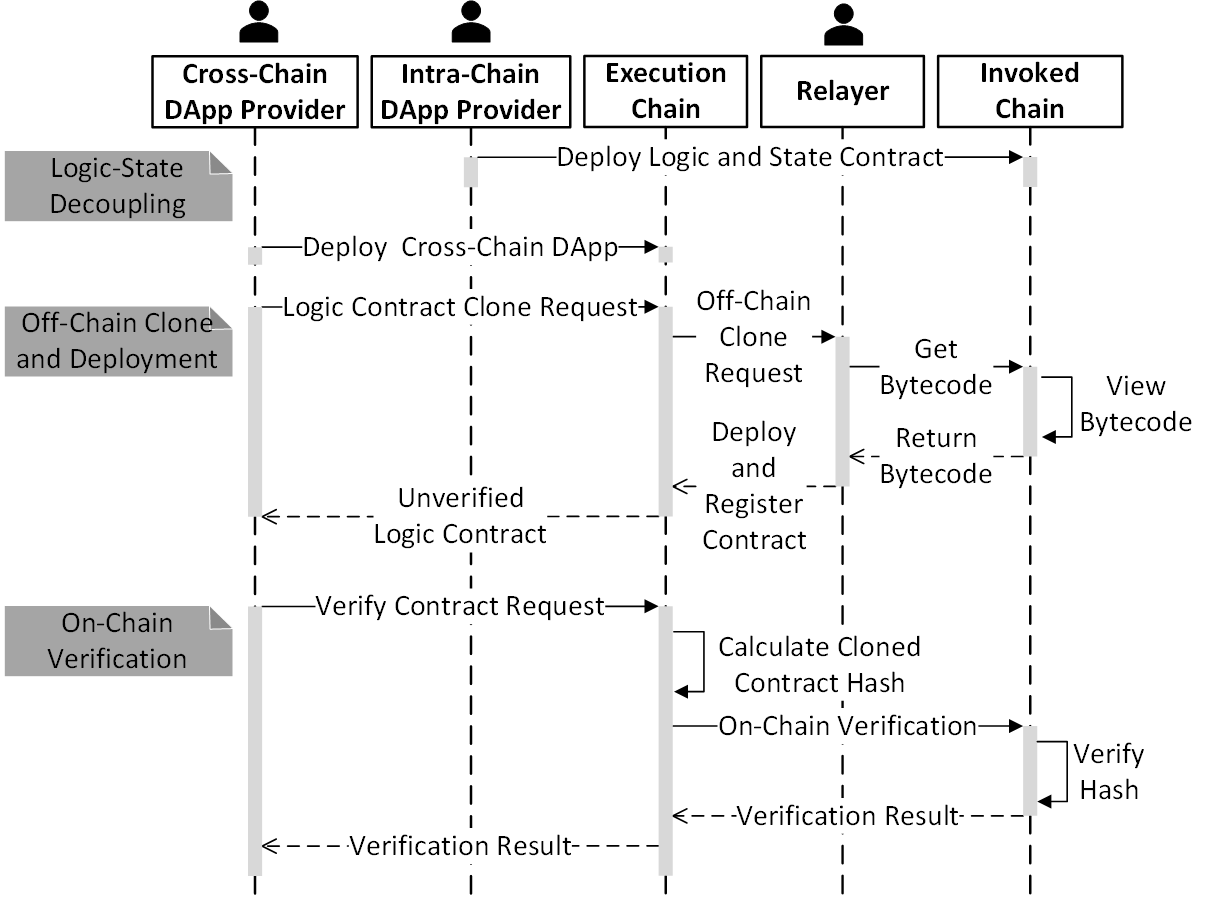}
    \vspace{-9pt}
    \caption{Hybrid Cross-Chain Smart Contract Deployment Protocol.}

    \label{MandV}
\end{figure} 

\subsection{Logic-State Decoupling}
\label{subsec:LSD}

To efficiently perform CCSCI, we need to deploy the logic of the invoked contracts to the same execution chain for efficient integrated execution. 
Existing contracts often contain both logic and state, and directly cloning such contracts would result in high gas costs and additional state storage overhead on the execution chain. 
Therefore, we design a set of Logic-State Decoupling (LSD) Guidelines to guide the developers to decouple existing contracts into logic execution contracts and state storage contracts. 
During cross-chain deployment, only the logic contracts need to be cloned, which reduces gas costs. 
Developers can follow these guidelines to develop new contracts with separated logic and state or upgrade existing contracts by decoupling them. 
We now provide a detailed explanation of the LSD Guidelines.
Moreover, a simple example illustrating the LSD is provided in Section \ref{codeex}.

\vspace{3pt}
\noindent
\textbf{State Contract.}
According to the LSD Guidelines, the decoupled state contract first includes all the \emph{variables} (representing states) from the original contract, as well as all the \emph{view functions} that read the contract's state. 
Since view functions are read-only and do not generate transactions, they do not affect cross-chain execution. 
Additionally, the state contract contains \emph{functions for locking and updating} the contract state, as cross-chain invocations require locking and updating states. 
The state contract should also contain \emph{functions that call the logic contract} in order to interact normally with the logic contract.
% To support normal on-chain calls, the state contract should also include functions that invoke the logic contract to execute the DApp's regular functionality.

\vspace{3pt}
\noindent
\textbf{Logic Contract.}
In the logic contract, no variables are stored. 
It only contains the functions that implement the original contract's \emph{execution logic}. 
These functions are called by the state contract to carry out the dApp's logic operations. 
When the state contract calls these functions, it passes all necessary state data (variables), and after the functions complete execution, the results are returned to the state contract. 
In our protocol, only the logic contract needs to be cloned, which reduces gas costs during cross-chain deployment.

\vspace{3pt}
\noindent
\emph{Remarks.}
Our protocol also supports existing smart contracts, even if they are not decoupled into logic and state. 
However, in such cases, the cross-chain deployment process will incur higher gas costs. 
% Furthermore, although the contracts migrated to the execution chain do not maintain contract states, they will be passively updated during each cross-chain call, leading to additional state storage costs.
In this case, the contracts deployed to the execution chain do not actively maintain their own state. 
Instead, they passively update their state during each cross-chain invocation. 
% This approach reduces the cost associated with active state updates.
The discussion related to developers' learning costs is given in Section \ref{disscussion}.

\vspace{3pt}
\subsubsection{Logic-State Decoupling Example}
\label{codeex}

We now give a sample of logic-state decoupling. In the following code~\ref{ex}, after applying logic-state decoupling, the original \texttt{Hotel} contract is split into two separate contracts: the logic contract \texttt{LHotel} and the state contract \texttt{SHotel}. The \texttt{LHotel} contract contains no state variables and only includes the \texttt{book()} function, which implements the hotel reservation functionality. Since there are no variables within the \texttt{LHotel} contract, the \texttt{book()} function must take all necessary parameters as inputs.

On the other hand, the \texttt{SHotel} contract retains all the variables from the original Hotel contract and introduces an additional address variable, \texttt{addr\_lhotel}, which records the address of the \texttt{LHotel} contract. In the \texttt{book()} function of the \texttt{SHotel} contract, no reservation logic is implemented directly; instead, it calls the \texttt{book()} function from \texttt{LHotel} using the \texttt{addr\_lhotel} parameter to execute the hotel reservation functionality.

By decoupling the logic and state in this way, only the \texttt{LHotel} contract needs to be cloned. 
% during logic updates. 
Since \texttt{LHotel} contains no state variables, this approach significantly reduces gas consumption during the cross-chain clone and deployment process.

\begin{lstlisting}[language=Solidity, caption={Pseudocode of Hotel Logic-State Decoupling}, label={ex}]
contract Hotel{
    int256 price;
    int256 remain;
    mapping (address => int256) accounts;
    function getPrice() public view returns(uint256); 
    function getRemain() public view returns(uint256); 
    function book(address user_addr, uint256 num) public returns(uint256); 
    function lockState(bytes[] memory args) public returns();
    function updateState(bytes[] memory args) public returns();

contract LHotel{
    function book(uint256 price, uint256 remain, uint256 num) public returns(uint256)
}

contract SHotel{
    int256 price;
    int256 remain;
    address addr_lhotel;
    uint256 lock_size;
    struct lock_bag;
    mapping (uint256 => lock_bag) lockpool;
    mapping (address => int256) accounts;
    function setLocksize(uint256) public returns();
    function getPrice() public view returns(uint256); 
    function getRemain() public view returns(uint256); 
    function lockState(bytes[] memory args) public returns();
    function updateState(bytes[] memory args) public returns();
    function book(address user_addr, uint256 num) public returns(uint256);

\end{lstlisting}

% \begin{figure}[t]
%     \centering
%     % 第一个子图
%     \begin{subfigure}[b]{0.525\textwidth}
%         \centering
%         \includegraphics[width=\textwidth]{Figures/migration2.png}  % 使用你的图片路径
%         \vspace{-10pt}
%         \caption{Off-chain clone and deployment}
%         \label{migration}
%     \end{subfigure}
%      % 第二个子图
%     \begin{subfigure}[b]{0.525\textwidth}
%         \centering
%         \includegraphics[width=\textwidth]{Figures/verification2.png}  % 使用你的图片路径
%         \vspace{-10pt}
%         \caption{On-Chain verification}
%         \label{verifaction}  % 子图标签，用于引用
%     \end{subfigure}
%     \vspace{-18pt}
%     \caption{Hybrid Cross-Chain Smart Contract Deployment Protocol.}
%     \label{MandV}  % 整体图的标签
% \end{figure}

\subsection{Off-Chain Clone and Deployment}
\label{subsec:migration}

Transmitting contract bytecode on-chain (via contract event) requires a significant amount of gas.
% ensures security, but it requires a significant amount of gas and incurs delays due to the need for blockchain consensus. 
To improve the efficiency of cross-chain logic deployment and reduce gas consumption, we propose \emph{transferring the contract bytecode via an off-chain solution}. 
The off-chain clone and deployment consists of two phases: the preparation phase and the clone phase. 
% In the following, we will explain the off-chain migration process through these two phases. 
The main process is shown in Figure ~\ref{MandV}.

\vspace{3pt}
\noindent
\textbf{Preparation.}
In this phase, the cross-chain dApp provider needs to obtain the call tree of smart contracts and determine which logic contracts need to be cloned.
\textcolor{red}{
Similar to existing works \cite{li2022jenga, robinson2021general, chen2024atomci}, this is achieved by first applying static analysis tools such as Slither~\cite{feist2019slither} to extract the call graph of the application.
External contract calls can be identified by parsing the call graph, where calls from one contract to another are marked. Static analysis tools can highlight these relationships, helping developers understand cross-contract dependencies. 
Therefore, from this call graph, the provider constructs the corresponding call tree, which captures the expected invocation paths during a single cross-chain transaction.
This process enables the identification of all logic contracts (a superset) that must be cloned to the invoked chain in order to ensure correct and atomic execution.
}
% can use static analysis tools, such as Slither~\cite{feist2019slither}, to obtain the call tree of smart contracts and determine which logic contracts need to be migrated. 
After this, the developer can choose a blockchain to deploy the cross-chain dApp. 
It is important to note that we offer developers a high degree of flexibility: 
They can select any blockchain according to their preference. 
For efficiency, the provider may pick a chain that already hosts some required logic contracts so that those do not need to be cloned again.
To reduce costs, they may also choose a blockchain that already hosts some required logic contracts,
\textcolor{red}{so that no additional relayer fees are incurred for cloning those contracts again. (Since cross-chain fees are ultimately paid by the provider and compensated to the relayers, this ultimately saves the provider's expenses.)}
Once the selection is made, the developer sends a transaction to invoke the bridging contract on the chosen chain.
The bridging contract then triggers an event to notify the relayers to initiate the cross-chain deployment.
% which triggers an event for contract migration. 
The event includes the ID of the invoked chain and the addresses of the logic contracts $Addr_{\text{L}}$ to be cloned on the invoked chain. 
This concludes the preparation phase.

{\textcolor{red}
\vspace{3pt}
\noindent
\emph{Remarks on Call Tree Analysis.}
Modern static analysis tools provide reliable extraction of contract call graphs and enable accurate construction of call trees. 
Even in the presence of misconfiguration or adversarial behavior by a dApp provider, the implications of an incorrect call tree are limited. 
Since the dApp provider is responsible for covering the gas costs of contract cloning, there is a built-in economic disincentive to over- or under-provision the call tree. 
An incorrect call tree may result in the cloning of unnecessary contracts on the execution chain, thereby increasing storage usage, but it does not introduce any security risk or compromise the atomicity of the cross-chain protocol.
Or, a missing logic clone would lead to a runtime exception on the execution chain, effectively aborting the cross-chain transaction. However, our protocol remains atomically safe (no partial commits occur on other chains, but the dApp’s operation would fail). 
The call tree is not stored on the execution chain.
Instead, it is constructed off-chain by the dApp provider during the preparation phase and is used solely as input to the bridging contract when initiating the cross-chain deployment.
Only the necessary information—such as the addresses of the logic contracts to be cloned and the identifier of the invoked chain—is packaged and passed as parameters to the bridging contract.
This design minimizes on-chain storage and computation overhead, while still providing relayers with the necessary context to coordinate cross-chain deployment.
}

\vspace{3pt}
\noindent
\textbf{Clone.}
After the event is triggered, relayers will detect the event and use the \texttt{getcode()} function from the bridging contract on the invoked chain to obtain the bytecode of the contract that needs to be cloned. 
This process is an \emph{off-chain read-only inquiry and, therefore, does not consume gas.}
Subsequently, the relayers will obtain the ABI file which defines the contract interface. 
The relayers will then deploy the contract to the execution chain. 
Once a relayer completes the redeployment, it registers the address of the cloned logic contract $Addr'_{\text{L}}$, through the bridging contract on the execution chain, marking the end of the clone phase.

\vspace{3pt}
\noindent
\emph{Remarks on Clone Reliability.}
For the cross-chain reliability of the protocol, multiple relayers perform the clone process after detecting the bridging contract's event. 
Therefore, even if some relayers do not respond to the event, other relayers will ultimately complete the contract clone and deployment.
Once one relayer completes the deployment, the bridging contract will trigger an event to stop the others, ensuring that the logic contract will not be deployed multiple times.
{\textcolor{red}
Furthermore, when a contract is cloned on the execution chain, the deployment transaction will invoke the \emph{register} function of the bridging contract deployed on the same chain. The bridging contract maintains a mapping from a service identifier (\texttt{string}) to the corresponding logic contract address.
Only the first registration for a given identifier will succeed. 
The deployment transaction will revert if the \emph{register} fails. This means the contract will not be created on the blockchain, and no bytecode will be stored. 
This mechanism prevents multiple cloning attempts, as any subsequent invocation of \emph{register} will fail, causing the entire cloning transaction to fail. As a result, only the first cloning attempt can successfully deploy the contract on the execution chain.
}

\textcolor{red}{
Moreover, if a malicious or overly eager relayer attempts to clone a logic contract without the presence of an event trigger, it must bear the gas cost independently.
Relayers are thus economically disincentivized from performing premature cloning, as no reimbursement is provided in the absence of an authorized event.
Even if such premature cloning occurs, it does not compromise system correctness; the resulting contracts are merely redundant and remain unused, leading at most to minor storage overhead.
}
\subsection{On-Chain Verification}
\label{subsec:verification}

Although off-chain clone and deployment can achieve efficient cross-chain logic deployment, it does not guarantee security as there may be malicious relayers present in the system.
A malicious relayer could potentially modify the contract bytecode or deploy a wrong contract. 
To ensure verifiability and security during the off-chain clone and deployment process,
% of the cloned and deployed logic contract, 
% we propose the on-chain cross-chain verification scheme to confirm the correctness of the cloned and deployed logic contract. 
we propose the on-chain cross-chain verification scheme to \emph{compare the cloned logic contract with the original contract and verify its correctness}.
The on-chain verification process is shown in Figure ~\ref{MandV}.

After the cross-chain deployment is completed, the cross-chain dApp provider can initiate cross-chain verification on the execution chain by calling the \texttt{Verification()} function of the bridging contract. 
The bridging contract will search the cloned contract bytecode of address $Addr'_{\text{L}}$, and calculate the hash of the bytecode. 
Then the bridging contract triggers an event includes the hash computation result. 
% The blockchain node will generate the Merkle proof~\cite{merkle1987} to verify the validity of the transaction.
{\textcolor{red}The relayers are responsible for transmitting this information (a cross-chain transaction) along with its \emph{receipt proof} and \emph{block header} to the invoked chain. The \emph{receipt proof} serves as a cryptographic proof that a specific event (e.g., a function execution or asset transfer) has indeed occurred on the source chain. It typically includes a Merkle proof~\cite{merkle1987}, which attests that the corresponding event log is embedded in a finalized block. 
The \emph{block header} provides the \emph{receiptsRoot}, which serves as the cryptographic commitment to the receipt set of the block and is used to verify the inclusion of the event log via the receipt proof.
This enables the invoked chain to verify the authenticity of the event without relying on trust assumptions about the relayer, thereby ensuring the integrity of cross-chain execution.}
In the invoked chain, the receipt proof with the cross-chain transaction is first verified to ensure the result has already reached consensus on the execution chain. 
The bridging contract then searches the corresponding local contract bytecode based on the address $Addr_{\text{L}}$, and calculates the hash of the bytecode.
Then, the bridging contract will verify whether it matches the hash transmitted across the chain. 
% calculates the hash of the local contract bytecode based on the address $Addr_{\text{L}}$ and verifies whether it matches the hash transmitted across the chain. 
The result of the verification is then returned.

{\textcolor{red}The bridging contract on the execution chain receives the verification result from the invoked chain.
To ensure authenticity, the execution chain first verifies that the provided \emph{block header} originates from the invoked chain and is finalized (e.g., via the bridging contract).
Then, using the accompanying \emph{receipt proof}, it reconstructs the Merkle path to the target receipt and verifies that the computed \emph{receiptsRoot} matches the one in the provided \emph{block header}.
This process guarantees the integrity and authenticity of the event data emitted on the invoked chain.
If the result confirms that the bytecode of the cloned contract on the execution chain matches the original, the bridging contract marks the cloned contract as verified.
Only verified contracts are permitted to participate in subsequent cross-chain invocations. Upon successful verification, the relayer responsible for the deployment is rewarded.
}
If the verification fails, the relayer will be penalized, and the off-chain clone and deployment process will be restart.

\vspace{3pt}
\noindent
\emph{Remarks.}
For the reliability of the cross-chain protocol, multiple relayers could transmit the same cross-chain transaction. 
However, the bridging contracts will deduplicate identical transactions from multiple relayers to avoid multiple executions on-chain.
This process also applies in the subsequent integrated execution protocol.
\section{Cross-Chain Smart Contract Integrated Execution Protocol}
\label{IntegratedExP}

\begin{figure}[t]
    \centering
    \includegraphics[width=0.48\textwidth]{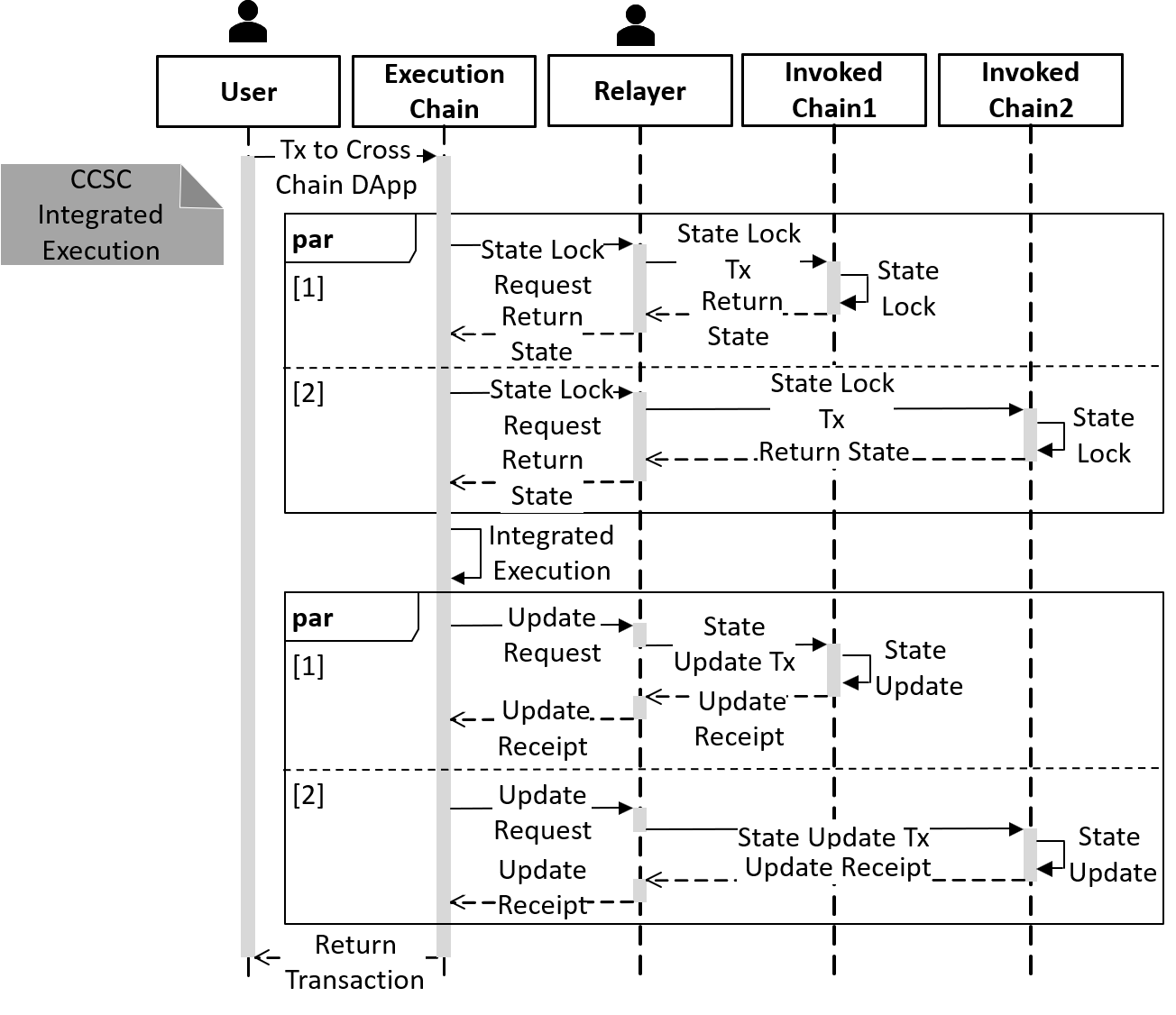}
    \vspace{-9pt}
    \caption{\texttt{IntegrateX}'s Cross-Chain Smart Contract Integrated Execution Protocol.}
    % An illustration of CCSCI process of sequential invocation and invocation in IntegrateX.}
    \label{ta}
     
\end{figure} 
% To enhance concurrency and reduce overhead
% of CCSCI while ensuring atomicity. We propose
% a Cross-Chain Smart Contract Integrated Execution Protocol. This protocol ensures high efficiency and atomicity throughout the entire call process through the Three-Phase Integrated Execution mechanism. In addition, we have incorporated finer transaction aggregation mechanisms and  granularity lock within the protocol to further enhance the efficiency of the \texttt{IntegrateX} system in complex cross-chain smart contract call scenarios.

\subsection{Atomic Integrated Execution}
\label{subsec:execution}

To efficiently and atomically execute complex CCSCI, we propose an atomic integrated execution scheme. 
As all the invoked logic has been migrated onto the execution chain, the atomic integrated execution does not need multiple rounds of cross-chain execution when handling CCSCI.
This enables all the logic to be executed within a single transaction, enhancing the efficiency of CCSCI. 
Moreover, to ensure overall atomicity for a series of CCSCI, we employ a state synchronization mechanism based on the Two-Phase Commit protocol~\cite{lampson1993twopc}. 
This process locks the relevant states across the involved chains, transmits the states to the chain responsible for integrated execution, and then returns the states to the respective chains to unlock and update them after the execution is completed. 
The entire atomic integrated execution consists Locking, Integrated Execution and Updating, which is shown in the Figure~\ref{ta}.

\vspace{3pt}
\noindent
\textbf{Locking. }
The Locking process locks all the required invoked contract states on the invoked chain.
The cross-chain dApp provider can obtain all the required state on the call tree beforehand via tools such as static analysis \cite{feist2019slither}.
Based on this information, a user sends a transaction via the cross-chain dApp to invoke the cross-chain dApp contract. 
% During this process, a user first sends a transaction to the cross-chain dApp contract. 
% the dApp contract can obtain all the required state on the call tree. 
The cross-chain dApp contract then calls the bridging contract to issue an event to lock the relevant states on the invoked chains.
% will issue an event to lock the relevant states on the invoked chain through the bridging contract. 
When relayers detect the event, it will transfer this message (i.e., cross-chain transaction with Merkle proof) to each invoked chain. 
The bridging contract on the invoked chain will verify the authenticity of the cross-chain though calculating the Merkle proof of the transaction and invoke the \texttt{LockState()} function of each invoked contract. 
Once the bridging contract has successfully locked the state and retrieved the required contract states, it will trigger an event to return the states. 
After the relayers detect the event, they will transmit these states via cross-chain transactions to the execution chain. 
After the execution chain's bridging contract verifies the authenticity of the transactions via the Merkle proof, it will return the states to the dApp.
Once all requested states are returned from individual invoked chains, the Locking process ends, and these states are used as inputs for the Integrated Execution.

\vspace{3pt}
\noindent
\textbf{Integrated Execution. }
The Integrate Execution process executes the entire CCSCI logic on the execution chain.
Cross-chain dApp contracts on the execution chain use the requested state values as inputs to perform the full call tree execution. 
Since all contracts required for the cross-chain invocation have completed logic migration and have been verified already, the integrated execution can be completed within a single transaction on the execution chain. 
The cross-chain dApp contract records the output results of each invoked contract during the Integrated Execution, allowing for state updates of the invoked contracts on other chains after the execution is completed. 
Once execution is complete, the Integrated Execution ends and transitions to the Updating process.

\vspace{3pt}
\noindent
\textbf{Updating. }
The Updating process unlocks and updates all the invoked contract states on the invoked chains.
After Integrated Execution is completed on the execution chain, the cross-chain dApp contract triggers an event to update the result via bridging contract. 
The relayers will distribute the result to the invoked chains. 
The bridging contract on each invoked chain verifies the Merkle proof of the cross-chain transaction and 
then invokes the \texttt{UpdateState()} function of each invoked contract, which will unlock and update the states of the invoked contracts.
\textcolor{red}{
It is important to note that in \texttt{IntegrateX}, the address of the bridging contract is a required parameter during the initialization of a state contract. In the \emph{state contract}, an \texttt{address} variable is used to store the address of the \emph{bridging contract}. Before executing any function related to state locking or updating, the contract verifies that the caller is the registered bridging contract. Only upon successful verification are the locking or updating operations permitted. This design allows the state contract to grant the bridging contract the permission to invoke specific internal functions, thereby enabling it to accept state locking and state updates initiated by the bridging contract.
}

\vspace{3pt}
\noindent
\emph{Remarks:}
\textbf{\emph{Rollback.}}
During the Atomic Integrated Execution, state rollback may occur due to failure in locking the state or execution failure. 
The invoked state might already be locked by another cross-chain invocation, which causes the failure to lock the invoked state. 
In this case, the bridging contract on the execution chain will initiate a new event to unlock all associated contracts that have already been locked and returned in this cross-chain invocation. 
For the contracts that have not yet completed the locking process, the event will cancel the lock attempt, thereby ensuring overall atomicity.
The execution might also fail, due to the reason such as insufficient gas fee or insufficient states.
% In the case of execution failure, 
% if the obtained states are insufficient to complete the cross-chain call, the call will fail. 
In this case, the bridging contract on the execution chain will discard the obtained states and initiate a cross-chain event to unlock all locked contracts, thus ensuring all invoked contracts are either fully locked or not locked at all.

\textcolor{red}{
For security concerns, within CCSCI, only the bridging contract is authorized (configured in the contract logic) to unlock the state contracts, while the relayer has no direct access to unlock any locked contract state. 
Furthermore, the relayer is incapable of forging an unlock request to the bridging contract, since any maliciously crafted message cannot pass the Merkle proof verification and will therefore be rejected during contract execution. 
This design ensures the security and integrity of CCSCI execution.
Besides, the rollback process does not involve any state updates. All pending state changes reside on the execution chain, and rollback merely discards the result of the integrated execution on the execution chain while releasing the locked states on the invoked chains. As such, \texttt{IntegrateX} does not  risk unintended consequences for other CCSCI.
}

% \noindent
\textbf{\emph{Timeout.}}
Additionally, to prevent the invoked contract state from being locked for an extended period, a design of timeout will be determined by the dApp developer within the application and managed by the execution chain. 
\textcolor{red}{
Timeout is a widely adopted mechanism in lock-based solutions~\cite{2019atomicBEswap,atomic-ibc}. In \texttt{IntegrateX}, both the bridging contract (set by our protocol) and the developer-defined smart contract can specify a maximum timeout. During execution, the effective timeout is determined by taking the minimum of these two values,  which prevents the dApp provider from setting an indefinite timeout period.
Moreover, timeout in blockchain systems is typically measured in terms of block depth rather than wall-clock time. Therefore, although the actual block time may vary across different blockchains, the maximum timeout in \texttt{IntegrateX} can be aligned approximately by using chain-specific block depths. 
For example, blockchains such as Ethereum\cite{eth} tend to have longer block intervals (e.g., ~12 seconds), so a timeout of 10 blocks would translate to roughly 2 minutes. In contrast, blockchains such as BNB Smart Chain~\cite{binance_whitepaper} may have shorter block times (e.g., 2-3 seconds), in which case a timeout of 10 blocks would correspond to a much shorter wall-clock duration. 
Moreover, the timeout is typically set to exceed the finality confirmation time of the underlying blockchain for safety.
% This approach ensures consistent logical behavior across heterogeneous chains. 
% Typically, the timeout duration is configured to exceed the finality confirmation time of the underlying blockchain.
}
For transactions that time out, such as when the locking is not completed or execution is not finished for an extended period, the execution chain will mark the cross-chain call as failed and send a cross-chain event to unlock the relevant contracts.
% That is because in a straightforward state-locking mechanism, 
% states locking can lead to a reduction in overall system call efficiency. To mitigate the impact of state locking and enhance the system's overall call efficiency and concurrency, we have designed a finer granularity lock. Once a contract state on the calling chain is requested, the requested state is locked, causing subsequent calls that attempt to change the state to fail due to the lock. To minimize such call conflicts and improve the overall efficiency of \texttt{IntegrateX}, we have implemented a finer granularity lock. finer granularity lock allows the state to be partially locked in a more granular manner, enabling it to be locked by multiple requests simultaneously, thereby increasing overall call efficiency. For contracts with frequent intra-chain calls, the finer granularity lock not only enhances cross-chain interoperability but also reduces intra-chain call blocking caused by cross-chain calls.

\textbf{\emph{Finality.}}
In \texttt{IntegrateX}, all cross-chain transactions must wait until the consensus on the initiating chain is finalized (or, for Nakamoto-type consensus~\cite{bitcoin}, highly likely to be finalized) before being committed to the execution chain or invoked chain, to ensure cross-chain security.
{\textcolor{red}
For different blockchains, \texttt{IntegrateX} allows the configuration of varying confirmation depths to ensure that the result of a transaction is sufficiently finalized. Specifically, it supports setting a block \emph{confirmation threshold} in the bridging contract, enabling tailored finality strategies for heterogeneous blockchains. 

To ensure that only finalized states are used on the execution/invoked chain, the bridging contract verifies the confirmation status of the submitted block header from the invoked/execution chain. 
It compares the submitted header with the latest block height of the invoked/execution chain and ensures that the block has reached the required confirmation depth. 
If this condition is not satisfied, the cross-chain message will be rejected (can be resend).
In addition, the bridging contract checks the validity of the message through the \emph{receipt proof}. Only when both the \emph{confirmation threshold} and \emph{receipt proof} verification succeed will the event be accepted and its associated state be used for further execution.
For different blockchains, the threshold setting is typically related to how long it takes for the blockchain itself to reach finality. 
Information regarding finality is usually defined in their project documentations. 
\texttt{IntegrateX} only needs to set the corresponding threshold in the bridging contract according to the definition in their documentations before connecting to their blockchains.
% For example, blockchains that use probabilistic finality mechanisms, such as those based on Nakamoto consensus~\cite{bitcoin}, require larger confirmation thresholds to reduce the risk of chain reorganization. In contrast, blockchains with deterministic finality, such as those based on Byzantine Fault Tolerant (BFT) consensus~\cite{castro1999pbft}, can adopt smaller thresholds due to their immediate and stable finality guarantees.
This flexible confirmation mechanism ensures that cross-chain execution is performed based on finalized and reliable states.
}

\subsection{Transaction Aggregation}
\label{subsec:aggregation}

% \begin{figure}[t]
%     \centering
%     \includegraphics[width=0.51\textwidth]{Figures/TAex.png}
%     \caption{An example for \texttt{IntegrateX}'s Transaction Aggregation mechanism.}
%     % An illustration of CCSCI process of sequential invocation and invocation in IntegrateX.}
%     \label{taex}

\begin{figure}[t]
    \centering
    \begin{subfigure}[b]{0.48\linewidth}
        \centering
        \includegraphics[width=\linewidth]{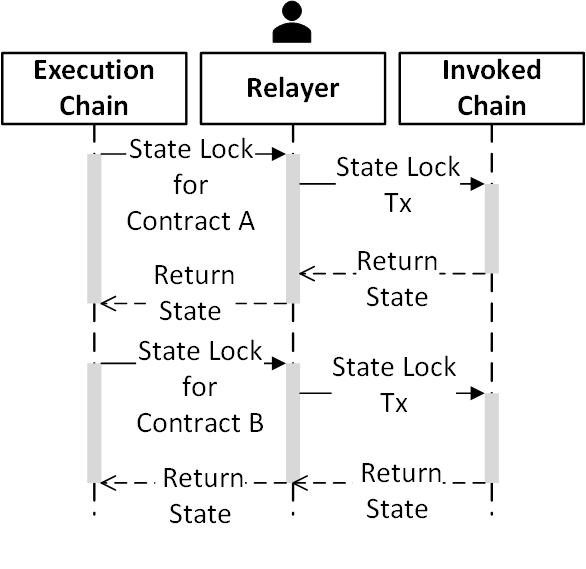}
        \vspace{-9pt}
        \caption{Invocation without TA}
        \label{tal}
    \end{subfigure}
    \hfill
    \begin{subfigure}[b]{0.48\linewidth}
        \centering
        \includegraphics[width=\linewidth]{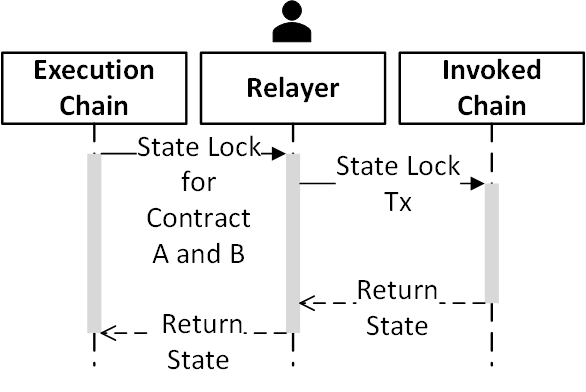}
        \vspace{-9pt}
        \caption{Invocation with TA}
        \label{tar}
    \end{subfigure}
    \vspace{-6pt}
    \caption{An example of \texttt{IntegrateX}'s Transaction Aggregation mechanism.}
    \label{taex}
\end{figure}

A large amount of cross-chain transactions for transmitting state incurs significant gas consumption. 
Handling CCSCI by sequential invocation may need to invoke contracts on the same chain in multiple rounds, which cause \emph{multiple rounds of cross-chain state transfers}. 
To address this issue, we design a transaction aggregation mechanism to reduce gas costs caused by multiple invocations of different contracts and state transfers on the same chain.
% To mitigate the gas consumption in this situation, we design the transaction aggregation mechanism to reduce gas consumption for cross-chain invocations involving contracts on the same chain for multiple times.

% Existing methods for ensuring atomicity in cross-chain invocation are sequential, meaning that calls are processed step by step, as shown in the Figure~\ref{ta}. Therefore, they need multiple rounds of cross-chain execution and cross-chain interaction in sequential order across several involved chains. Therefore, when invoking multiple contract states on the same invoked chain, it is necessary to access that chain multiple times. However, 
Our protocol locks all required states simultaneously, allowing multiple states on the same chain to be locked together, even when the calls are non-contiguous. 
The transaction aggregation mechanism combines all state requests on the same chain into a single transaction, reducing the number of cross-chain transactions. 
Similarly, during state updates, the transaction aggregation mechanism reduces the number of update requests. 
Therefore, for cross-chain calls involving multiple contracts on the same chain, this approach ensures that the number of cross-chain transactions is equal to the number of invoked chains (rather than the number of contracts), thereby reducing gas consumption.

\textcolor{red}{
Figure~\ref{taex} shows an example that the execution chain needs to lock the state of both \emph{Contract A} and \emph{Contract B} on the invoked chain. Sequential invocation needs to invoke \emph{Contract A} and \emph{Contract B} in multiple rounds, as illustrated in Figure~\ref{tal}. 
By using transaction aggregation mechanism, the execution chain only needs to send a single transaction to lock the state of both \emph{Contract A} and \emph{Contract B}, as shown in Figure~\ref{tar}. 
Through this mechanism, the execution chain can avoid multiple rounds of cross-chain messaging, improving efficiency. Furthermore, the reduction in the number of transactions also leads to lower gas consumption.
}
\subsection{Fine-Grained State Lock}
\label{subsec:lock}

During the atomic integrate execution process, we need to lock contract states to ensure atomicity. 
A simple approach is to either lock the entire state of the invoked contract or lock individual states being invoked~\cite{chen2024atomci, robinson2021general}. 
However, these state-locking mechanism reduces transaction concurrency.
Because once a state is locked, any subsequent transactions related to that state will fail until the state is unlocked. 
Therefore, we establish a set of guidelines to guide developers in decomposing certain states that can be split into finer granularity, and allowing \emph{partial state locking}. 
Unlike existing protocols that require locking the entire state, our fine-grained state lock mechanism locks only partial of a state at a fine-grained level. This approach enhances concurrency by reducing unnecessary state locking.

In EVM-compatible blockchains, Solidity is the smart contract language. 
In Solidity, the state of a contract is typically represented by \emph{variables}. 
Various types of variables are used in Solidity, such as \texttt{uint}, \texttt{address}, and \texttt{boolean}. 
% We observe that \texttt{uint} variables are the most widely applied. 
We find that the \texttt{uint} variable is widely used and can be decomposed.
Based on this observation, we design a fine-grained state lock specifically for \texttt{uint} variables and develop a lock pool mechanism. 
The lock pool is a structure where, during the use of the fine-grained state lock, part of the state is locked within this structure until execution is completed, while the unlocked portion of this state remains accessible, thereby enhancing the concurrency of the application.

% For variables directly used based on user input, the exact value of the required state can be precisely determined. 
For variables that can be directly derived from transaction inputs, the exact value of the required state can be precisely determined.
A fine-grained state lock can be applied to accurately lock only the relevant portion of such state.
For states that are dynamically used during execution, their exact values cannot be determined at the beginning. 
We allow dApp developers to lock these states in fixed-size increments based on their needs.
% For states dynamically used during execution, which can not be precisely determined at the beginning, the fine-grained state lock will lock them in fixed-size increments. 

{\textcolor{red}
For example, as shown in Listing~\ref{ex} in Section~\ref{codeex}, users can book hotel rooms through a smart contract. For simplicity, we assume all rooms are of the same type, and the \emph{remain} variable represents the number of available rooms.
Without fine-grained state locking, a booking attempt by one user would lock the entire contract state, preventing concurrent transactions.
In the \emph{SHotel} contract, a fine-grained locking mechanism is implemented using a \emph{lock\_bag} structure, which allows partial locking of the \emph{remain} variable. This enables other transactions to access and interact with the remaining state.
During the locking period, a \emph{lockpool} temporarily holds the locked portion of the state.
Developers can configure the \emph{lock\_size} parameter to determine how much of the state is locked per transaction, helping to prevent failures caused by insufficient locked state.
}

\vspace{3pt}
\noindent
\emph{Remarks.}
We focus on the flexibility of this mechanism, allowing developers to choose whether to implement the finer granularity locking mechanism in their dApp and to set the lock granularity based on the specific needs of the dApp. 
Moreover, developers can set the fixed size of the fine-grained state lock based on their preferences, tailored to the specific use case of the application.
More discussion related to developers' learning costs is given in Section \ref{disscussion}.
% 在Hybrid Cross-Chain Smart Contract Deployment Protocol 协议中，relayer对需要被跨链调用intra-chain DApp的逻辑合约进行了迁移。relayer在bridging合约中通过已经部署的合约地址获得其字节码，合约的 API 文件用来定义与合约交互的接口，通常在智能合约部署时就会公开，因此可以直接获取。通过 API 文件以及链上合约的字节码，relayer可以重新在其他区块链上部署该合约。在IntegratedX系统中，Hybrid Cross-Chain Smart Contract Deployment Protocol 协议内的任意relayer都可以进行迁移操作，然而这些relayer并不一定是可信任的，这就导致了可能有恶意的relayer在迁移过程中故意更改字节码造成安全性漏洞，因此我们通过on-chain verification进行链上合约验证来保证字节码没有被更改。
% 我们设计使用bridging合约比较迁移后合约的字节码哈希值和原合约的字节码哈希值。合约字节码哈希值可以通过智能合约的链内调用获得，并且将会由relayer传输到intra-chain DApp所在的链上，由于存在至少一个可以运行的relayer，该哈希值最终会被传输到目标链上并由目标链进行验证并将验证结果回传。由于区块链数据的不可篡改性，使用链上智能合约验证可以保证验证结果无法被更改。只有通过验证的合约才会完成完整的登记流程并可以被跨链应用链内调用。由于区块链数据具有可追溯性，因此对于验证通过的合约，进行该次迁移的relayer将会获得奖励，如果验证失败，将由新的relayer进行迁移，并对迁移失败的relayre进行惩罚。惩罚措施包含建立黑名单机制，一定次数的验证失败将无法进行迁移且此次迁移的费用将会由relayer承担。通过这样的on-chain verification，可以保证逻辑合约迁移后的正确性。由于至少存在一个relayer可以正常工作，因此该逻辑合约最后一定会被成功迁移。
\section{Security Analysis}\label{security_analysis}
% In this section, we analyzed how \texttt{IntegrateX} can safely and efficiently perform complex cross-chain calls.

\subsection{Security in Hybrid Cross-Chain Smart Contract Deployment Protocol}

\begin{theorem}
The Hybrid Cross-Chain Smart Contract Deployment Protocol ensures reliability, verifiability, and consistency, if the proportion of Byzantine nodes in each blockchain is less than its fault tolerance threshold, and at least one functional relayer is present.
% As long as the number of Byzantine nodes in each blockchain remains within the Byzantine fault tolerance limits, and at least one functional relayer is present, the reliability, verifiability, and consistency of the Hybrid Cross-Chain Smart Contract Deployment Protocol can thus be assured.
\end{theorem}

\begin{proof}
\noindent
\textbf{Reliability. }
The Hybrid Cross-Chain Smart Contract Deployment Protocol ensures reliability, meaning that when an execution chain issues a request for cross-chain deployment of a contract, the requested contract will eventually be deployed to the execution chain. 
Within the Hybrid Cross-Chain Smart Contract Deployment Protocol, multiple relayers listen for such requests. Even if malicious relayers intentionally ignore the requests, assuming that at least one functional relayer exists, this relayer will ultimately handle the logic clone and deployment of the requested contract. Thus, even in the worst-case scenario, the protocol ensures that the contract will be successfully deployed to the execution chain.

\vspace{3pt}
\noindent
\textbf{Verifiability. }
The Hybrid Cross-Chain Smart Contract Deployment Protocol ensures verifiability, meaning that both the execution chain and the invoked chain are able to verify the cross-chain transactions transmitted by the relayers. 
Additionally, both chains can verify the hash values of the contract bytecode before and after the cross-chain clone and deployment to ensure that the contract has been deployed correctly. 

In the Hybrid Cross-Chain Smart Contract Deployment Protocol, multiple relayers listen to the request and relay the messages. 
% However, relayers are not responsible for verifying the cross-chain transactions. 
% Instead, 
The invoked chain (bridging contract) validates the authenticity of the transactions using the Merkle proof attached with the cross-chain transactions, thereby preventing the relayers from altering the cross-chain data. 
By parsing the transactions, the invoked chain can obtain the bytecode hash of the cloned contract and compare it with the bytecode hash of the original contract on the chain to ensure the correctness of the cross-chain deployment. 
Additionally, the nonce value associated with each transaction prevents malicious replay attacks.

{\textcolor{red}
It is worth noting that the concept of a relayer is not originally proposed by \texttt{IntegrateX}, nor is the assumption of at least one honest relayer unique to this system. In fact, all existing cross-chain protocols rely on the presence of at least one genuinely honest and non-compromised relayer~\cite{robinson2021general,atomic-ibc,chen2024atomci,liu2021hyperservice,hu2024ivyredaction}. This relayer must not be forgeable or impersonated by an adversary; otherwise, the threat model would be fundamentally violated and the protocol's security guarantees would no longer hold.
During cross-chain communication, the relayer solely acts as a message carrier. It listens to on-chain events and forwards the corresponding message along with a \emph{receipt proof} (mentioned in Section \ref{subsec:verification}), which attests to the inclusion of the event in the block header of the source chain. Upon receiving the message, the invoked chain verifies the \emph{receipt proof} to ensure the message was indeed generated by the execution chain. Only after successful verification does the \emph{bridging contract} on the invoked chain reconstruct and execute the intended contract call.
This verification mechanism is symmetric and also applies to messages sent from the invoked chain back to the execution chain, thereby ensuring the integrity of cross-chain communication regardless of relayer trustworthiness.
Therefore, when inconsistencies arise due to a relayer maliciously modifying the transmitted message, the tampered data will be rejected by the invoked chain as it fails the on-chain proof verification. Only messages that pass the verification will be accepted.

Furthermore, in the case of blockchain forks leading to conflicting relayer messages, \texttt{IntegrateX} defers execution until one fork becomes finalized (mentioned in Section~\ref{subsec:execution}). Once finality is reached, the corresponding transaction results are adopted for further processing.
}

\vspace{3pt}
\noindent
\textbf{Consistency. }
The Hybrid Cross-Chain Smart Contract Deployment Protocol ensures consistency, meaning that during the off-chain clone and deployment as well as the on-chain verification process, both the execution chain and the invoked chain reach agreement on the outcome of the cross-chain requests.
% meaning that the agreement between the execution chain and the invoked chain on the outcome of the cross-chain request during off-chain clone and deployment and on-chain verification. 
In the Hybrid Cross-Chain Smart Contract Deployment Protocol, the proportion of malicious nodes on both the execution chain and the invoked chain remains within the fault tolerance threshold.
Moreover, the protocol requires to wait until consensus on one chain is finalized (or highly likely to be finalized) before committing the cross-chain transaction to another chain. 
% both the execution chain and the invoked chain maintain within the Byzantine fault tolerance limits. 
As a result, even in the presence of Byzantine nodes, each chain can still achieve consensus on cross-chain transactions and finalize them, ensuring that consistency is not undermined by malicious nodes. 
This guarantees that the outcomes of cross-chain requests remain consistent.
\end{proof}

\subsection{Security in Cross-Chain Smart Contract Integrated Execution Protocol}
%Cross-Chain Smart Contract Integrated Execution Protocol实现了跨链调用的原子执行，IntegrateX系统通过类似于 2PC 的状态同步机制对invoked的合约进行状态锁定和更新，在cross-chain应用中当用户在execution chain上的调用cross-chain DApp合约后，bridging合约将向所有invoked chain上的invoked合约发送状态请求，如果任何一个invoked合约未能锁定状态，将会返回失败信息，bridging合约将会取消本次调用并且发送跨链请求解锁所有其他合约，被锁定的合约将被解锁，而当尚未完成锁定的合约先接收到了解锁请求，将会在忽略后续接收到的该次锁定请求，以此保证合约状态锁定的一致性。
% 当所有被请求的合约都成功锁定状态后，cross-chain DApp合约将在execution chain上进行集成执行并将结果返回给所有的invoked合约，invoked 合约的新的状态将会首先返回给invoked chain上的bridging合约，并由bridging合约返回状态更新成功信息，如果任意一条invoked chain上的状态更新失败，所有链上的bridging合约都会丢弃更新的状态，此次跨链调用将失败，只有所有invoked chain上都完成了状态更新，cross-chain DApp合约才会输出最终结果，invoked合约解锁状态并从本链的bridging合约进行状态更新，标志着此次调用成功，由此保证完整的跨链调用原子性。
\begin{theorem}
The Hybrid Cross-Chain Smart Contract Integrated Execution Protocol ensures overall atomicity, reliability, verifiability, and consistency, if the proportion of Byzantine nodes in each blockchain is less than its fault tolerance threshold, and at least one functional relayer is present.
% As long as the number of Byzantine nodes in each blockchain remains within the Byzantine fault tolerance limits, and at least one functional relayer is present, the atomicity, reliability, verifiability, and consistency of the Cross-Chain Smart Contract Integrated Execution Protocol can thus be assured.
\end{theorem}

\begin{proof}
\noindent
\textbf{Overall Atomicity. }
The Cross-Chain Smart Contract Integrated Execution Protocol guarantees overall atomicity, which means that in the selected CCSCI process, state changes on both the execution chain and the invoked chain either all succeed or all fail, preventing any situation where one chain's state changes while the other does not. 
We employ the atomic integrated execution mechanism, similar to the 2PC scheme. 
For one cross-chain dApp, during this process, all the states required by the invocation on the invoked chains will be locked. 
If any contract has already been locked by another invocation, this invocation will fail, and all other invoked contracts will be unlocked. 
If the execution chain obtains all the necessary states but the execution fails due to insufficient gas or other reasons, the execution will be aborted, and all related locked contracts will be unlocked without any state changes. 
Furthermore, the protocol incorporates a timeout scheme: 
If any of the invoked chains fails to return the required state within the specified time frame by dApp, or if the execution transaction on the execution chain fails to complete within the specified time limit, the execution chain will abort the invocation and unlock all related contract states, ignoring any subsequent state responses from the invoked chains. 
% In cases where an invoked chain receives a lock request after receiving an unlock request, it will also ignore the lock request, ensuring that all invoked contract states are either locked or unlocked simultaneously. 
As a result, only when the execution chain has successfully acquired all required states and completed execution will it issue state updates to all related invoked chains, thereby ensuring the atomicity of the entire CCSCI process.

\textcolor{red}{
It is important to note that, similar to existing cross-chain solutions, \texttt{IntegrateX} incorporates an acknowledgment (receipt) mechanism for cross-chain messaging~\cite{ye2020bitxhub,atomic-ibc}. In the \emph{Updating} phase of integrated execution, the execution chain receives receipts from all invoked chains. The final result is output on the execution chain only after all receipts confirm that the state updates have been successfully completed on the respective chains. This mechanism ensures that the execution chain is informed whether the updating states transaction has been successfully executed.
Only upon successful execution will the system proceed to the next step. Any failure in the cross-chain execution triggers a full rollback of the transaction, thereby preserving the overall atomicity of \texttt{IntegrateX}.
As a result, even if an update operation on the invoked chain fails (which barely happens), the entire transaction will be rolled back, maintaining the system’s atomicity guarantees.
}

\vspace{3pt}
\noindent
\textbf{Reliability. }
The Cross-Chain Smart Contract Integrated Execution Protocol ensures reliability, which means that when an execution chain initiates a CCSCI request, the invoked chain will eventually receive the cross-chain transaction, and the state returned by the invoked chain will likewise be received by the execution chain. 
In the Cross-Chain Smart Contract Integrated Execution Protocol, multiple relayers monitor CCSCI requests. 
Even in the presence of malicious relayers who deliberately fail to respond to the request, the assumption of at least one functional relayer ensures that, in the worst-case scenario, this relayer will relay the cross-chain transaction to the invoked chain, ensuring that the transaction is eventually received. 
Similarly, even in the worst-case scenario, at least one relayer will transmit the state returned by the invoked chain back to the execution chain, thereby guaranteeing the reliability of the CCSCI process.

\vspace{3pt}
\noindent
\textbf{Verifiability. }
The Cross-Chain Smart Contract Integrated Execution Protocol ensures verifiability, which means the ability of the invoked chain to verify the authenticity of cross-chain transactions transmitted by relayers from the execution chain, while the execution chain can also verify the transactions returned by the invoked chain. 
In this protocol, multiple relayers listen for the request and relay messages.
% , but relayers themselves are not responsible for verifying the cross-chain transactions. 
Both the execution chain and the invoked chain can independently validate the authenticity of the cross-chain transactions using the Merkle proof attached with the transactions. 
Additionally, the use of transaction nonce values prevents malicious replay attacks, ensuring the integrity of the cross-chain interaction.

\vspace{3pt}
\noindent
\textbf{Consistency. }
The Cross-Chain Smart Contract Integrated Execution Protocol guarantees consistency, ensuring that both the execution chain and the invoked chain agree on the result of the requested operation during a cross-chain smart contract invocation. 
In this protocol, the proportion of malicious nodes on both chains remains within the fault tolerance threshold, allowing consensus to be achieved even if there exist Byzantine nodes. 
Additionally, the protocol requires waiting until consensus on one chain has been finalized before committing the cross-chain transaction to the other chain.
Therefore, cross-chain transactions on the blockchain cannot be maliciously altered, ensuring that both the execution chain and the invoked chain reach a unified agreement on the outcome of cross-chain operations.
\end{proof}

\section{Implementation and Evaluation}\label{evaluation}

% In this section, we implement \texttt{IntegrateX} protocol with local servers and conduct comprehensive tests to evaluate the performance of the protocols in terms of run time, latency, and gas consumption.

\subsection{Implementation}
% The \texttt{IntegrateX} system incorporates on-chain components on both the invoked chain and the execution chain. 
We implement a prototype of \texttt{IntegrateX} based on a cross-chain communication project Bitxhub \cite{ye2020bitxhub} (its transport layer is similar to IBC). 
The relayers are implemented in Golang, while the bridging contracts are implemented using Solidity.
{\textcolor{red}For the underlying blockchain, we do not rely on a custom-built chain. Instead, we leverage the existing \texttt{go-ethereum} client to implement several EVM-compatible blockchains in Golang, without making any modifications to the underlying blockchain infrastructure.}
For performance comparison, we also implement a baseline CCSCI protocol GPACT~\cite{robinson2021general}. 
{\textcolor{red}
As mentioned in the paper, GPACT is one of the most advanced existing protocols that can guarantee overall atomicity for complex CCSCI. 
However, GPACT has limitations in efficiency, as it requires multiple rounds of cross-chain execution and message exchange.
}
Furthermore, we use Solidity to implement the Train-and-Hotel example mentioned in the paper, along with smart contracts in various other scenarios, to demonstrate the \texttt{IntegrateX}'s performance in real use cases and under different conditions.

{\textcolor{red}
For the ease of implementation and experimentation, the underlying blockchain employs a Proof-of-Authority (PoA) mechanism for leader selection, and achieves consensus through Practical Byzantine Fault Tolerance (PBFT). PoA is a consensus protocol in which a limited set of pre-approved validators, referred to as authorities, are responsible for block production. 

To support our proposed protocols, we implement a bridging contract on the blockchain. The protocol requires no modification to the underlying blockchain infrastructure; it only involves deploying these bridging contracts.
We implement the functions \emph{regServer} and \emph{regState} to register the logic contract and the state contract, respectively.
During the on-chain verification process, the \emph{compareBytes} function in the bridging contract is used to compare the bytecode of the cloned contract with that of the original contract on the invoked chain, ensuring that the contract has been correctly cloned.
In the Cross-Chain Smart Contract Integrated Execution Protocol, the bridging contract is also responsible for issuing the CCSCI transaction. To this end, we implement the \emph{lockState} and \emph{updateState} functions to lock and update the contract state on the invoked chain.
Importantly, this integration is lightweight and non-intrusive, requiring no changes to the underlying blockchain beyond smart contract deployment.

Furthermore, \texttt{IntegrateX} supports cross-chain interoperability via a relayer built upon BitXHub’s cross-chain communication framework. The relayer includes a receipt mechanism to ensure the reliability and traceability of message delivery. 
Unlike traditional cross-chain systems that rely on trusted relayers for message validation, the relayer in \texttt{IntegrateX} is entirely trustless. Its sole responsibility is to relay raw cross-chain data—such as events, receipts, and Merkle proofs—between chains. All critical verification, including block header validation, Merkle proof checking, and event authenticity verification, is executed on-chain through smart contracts.
This design eliminates the need to trust any off-chain component and ensures that incorrect or forged messages submitted by a malicious relayer will be rejected by the verification logic on the execution chain. Consequently, the relayer can be treated as a permissionless data courier, significantly enhancing the security and decentralization of the system.
}

% utilized the open-source project Bitxhub \cite{ye2020bitxhub} to implement the cross-chain communication component of the \texttt{IntegrateX} system. 
% % In BitXHub, cross-chain information is transmitted by the relayer through the cross-chain light client. 
% We implemented the light client in Golang. We implemented bridging contracts and various dApp applications using Solidity, including an application that increases call depth, the train-hotel application, and others. Additionally, we implemented versions of the aforementioned applications under GPACT~\cite{robinson2021general} using Solidity. In GPACT, CCSCI requires multiple rounds of execution. By comparing the latency and gas consumption of the IntegrateX system and GPACT under CCSCI, we can better demonstrate the efficiency of the IntegrateX system in CCSCI. 

\subsection{Experimental Setup}

We test the performance of \texttt{IntegrateX} on multiple servers.
Each server is equipped with an Intel® Core(TM) i5-10400F CPU @ 2.90 GHz and 15.5 GB RAM running 64-bit Ubuntu 22.04 LTS. 
% We build the underlying EVM-compatible blockchain in Golang. 
We deploy 4 relayers via Docker containers on multiple servers.
We also deploy 3 blockchains, each with 4 nodes.
\textcolor{red}{
To decide block time, we notice that existing well-known EVM-compatible blockchains (e.g., BNB Smart Chain~\cite{binance_whitepaper}, Polygon~\cite{keil2000polygon}, Optimism~\cite{carver2010optimism}, Ethereum~\cite{eth}) typically adopt block times ranging from 2 to 12 seconds. As a result, the default block time in the experiments is set to be 5 seconds to maintain generality while ensuring a reasonable balance between performance and realism. 
% we chose a 5-second block time as a representative median value.
}
% We deploy serveral Docker containers on multiple servers to run the system, including 4 relayers and 3 cross-chain light clients. 
We set each block to contain up to 4096 transactions.
% , and we assume that the block gas limit is not reached.
Additionally, the gas calculation method in the experiments is the same as that used in Ethereum. 
For every experiment, we select Blockchain 1 (bc1) as the execution chain in the \texttt{IntegrateX} system, with the other chains as invoked chains.
% Detailed descriptions of the experimental setup will be provided before presenting the performance evaluation results.

\begin{figure}[t]
    \centering
    % 第一个子图
    \begin{subfigure}[b]{0.22\textwidth}
        \centering
        \includegraphics[width=\textwidth]{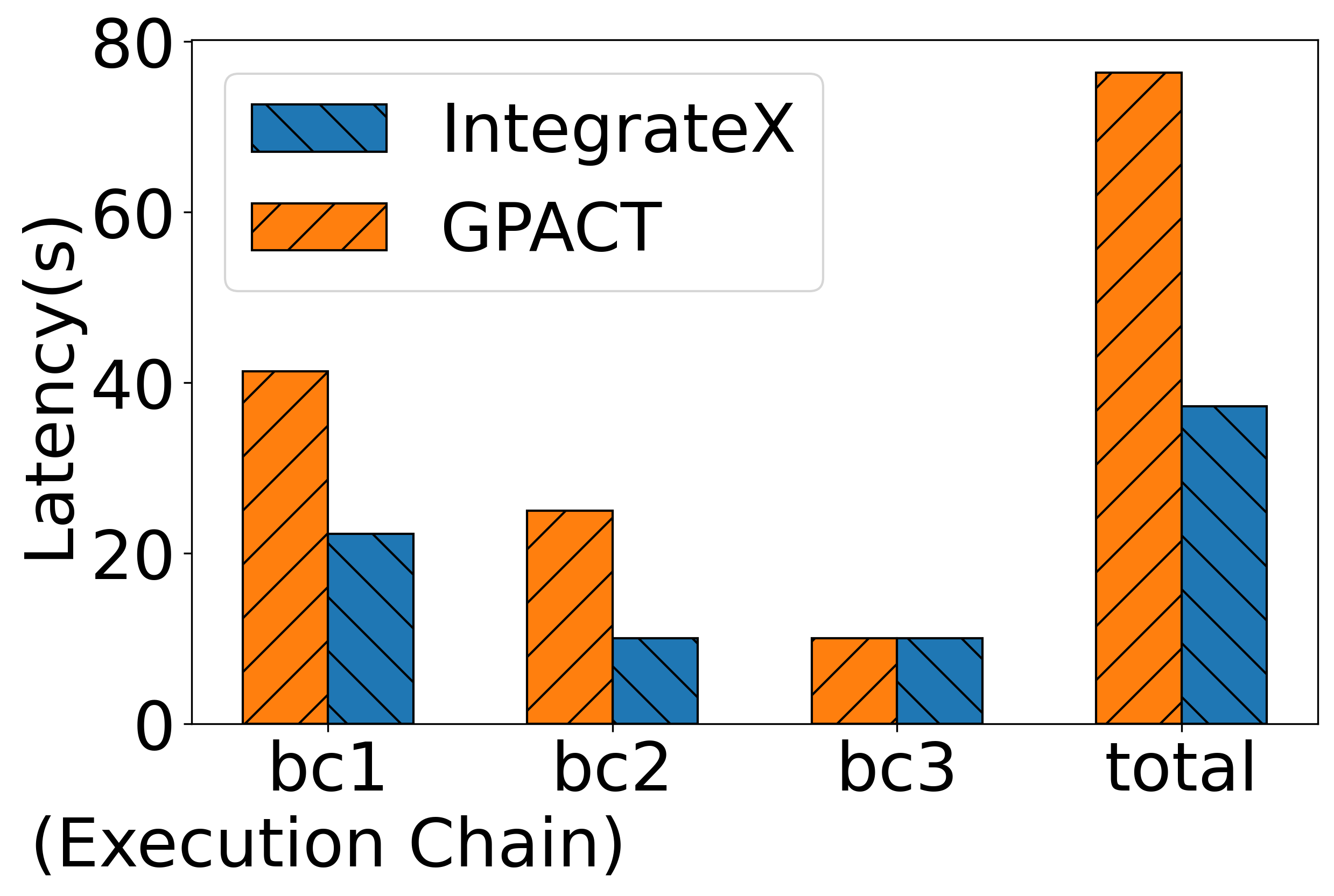}  % 使用你的图片路径
        \caption{Latency}
        \label{ixtime}
    \end{subfigure}
    % \hspace{3pt}
    \hfill
     % 第二个子图
    \begin{subfigure}[b]{0.22\textwidth}
        \centering
        \includegraphics[width=\textwidth]{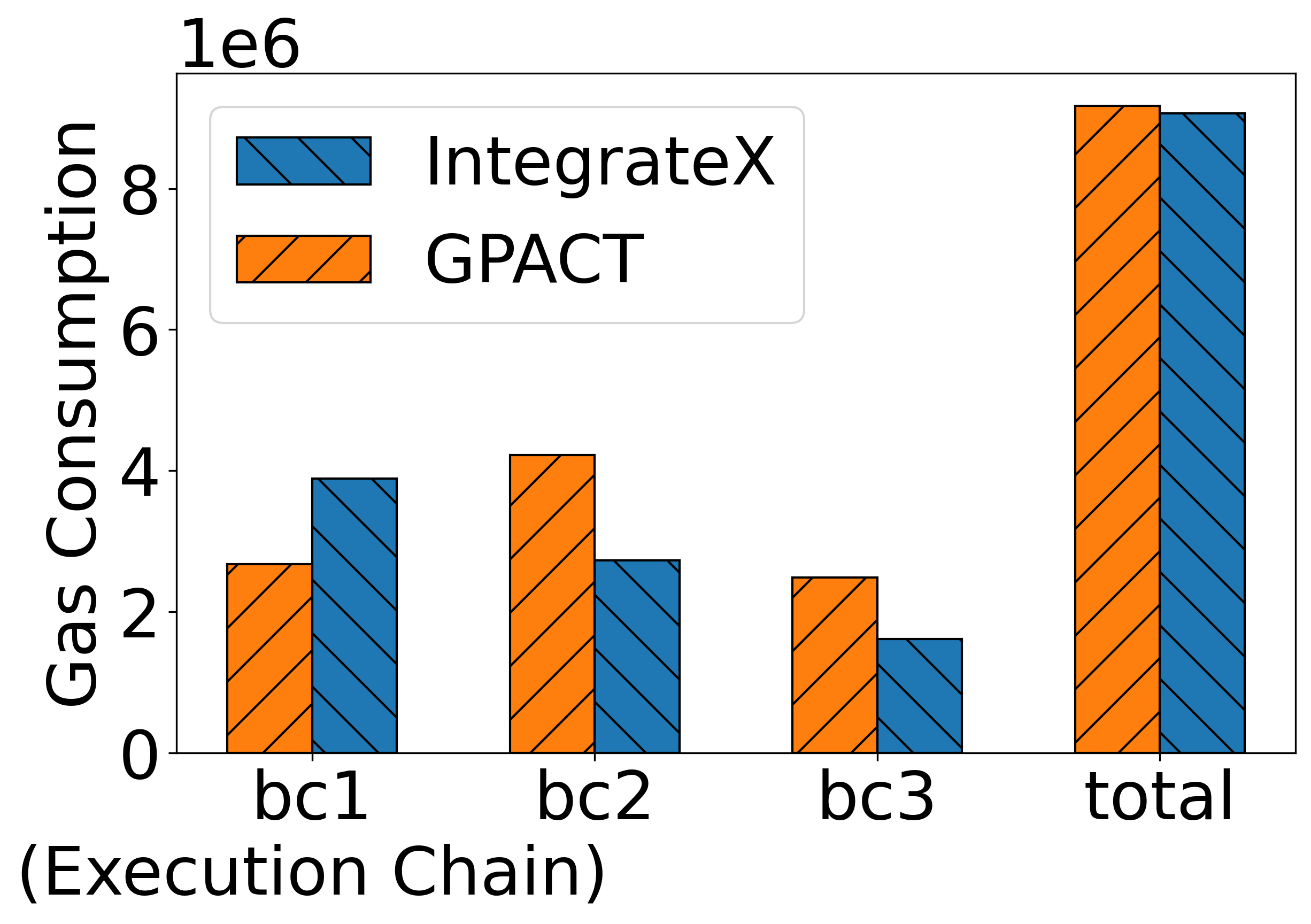}  % 使用你的图片路径
        \caption{Gas consumption}
        \label{ixgas}  % 子图标签，用于引用
    \end{subfigure}
    \vspace{-6pt}
    \caption{Latency and gas consumption in the Train-and-Hotel use case.}
    % TA: transaction aggregation.}
    \label{integratex}  % 整体图的标签
    \vspace{-6pt}
\end{figure}

\subsection{Experimental Results}
\subsubsection{Performance under Train-and-Hotel Use Case}
% In this experiment, 
We now compare the latency and gas consumption during the Train-and-Hotel CCSCI process for \texttt{IntegrateX} and GPACT. 
% The results are shown in Figure~\ref{integratex}.

\texttt{IntegrateX} decreases total latency by 51.2\% compared to GPACT, as shown in Figure~\ref{ixtime},. 
On bc1 (execution chain), the latency of \texttt{IntegrateX} is reduced by 46.2\% compared to GPACT.
The reason is that, due to the integrated execution, \texttt{IntegrateX} could integrate and execute all the related execution logic within one block, avoiding multiple rounds of cross-chain execution during CCSCI.
On bc2, \texttt{IntegrateX} also reduces the latency by 60\% compared to GPACT.
% This is because that the Train Contract is invoked multiple times, 
This is because the train contract is invoked multiple times, requiring multiple rounds of state transfers in GPACT, whereas in \texttt{IntegrateX}, only a single round of state transfer is needed.
% Train-and-Hotel use case invokes the Train Contract state on bc2 multiple times, the transaction aggregation mechanism in \texttt{IntegrateX} locks all the required states in a single transaction, further enhancing efficiency. 
% Overall, in the Train-and-Hotel use case, compared to GPACT, \texttt{IntegrateX} achieves a 51.2\% improvement in efficiency with nearly the same gas consumption.

The results in Figure \ref{ixgas} show that the total gas consumption of \texttt{IntegrateX} and GPACT is nearly identical in the Train-and-Hotel use case. 
On bc1, the gas consumption in \texttt{IntegrateX} is larger than GPACT (41.9\%).
This is mainly because in \texttt{IntegrateX}, the execution chain (bc1) handles all the logic during CCSCI, leading to a higher gas.
However, on the invoked chains (bc2 and bc3), \texttt{IntegrateX} achieves lower gas cost compared to GPACT (35.4\%, 35.2\%, respectively).
The main reason is that in \texttt{IntegrateX}, the invoked chains only need to lock and update states without executing any logic. 

\begin{figure}[t]
    \centering
    \includegraphics[width=0.22\textwidth]{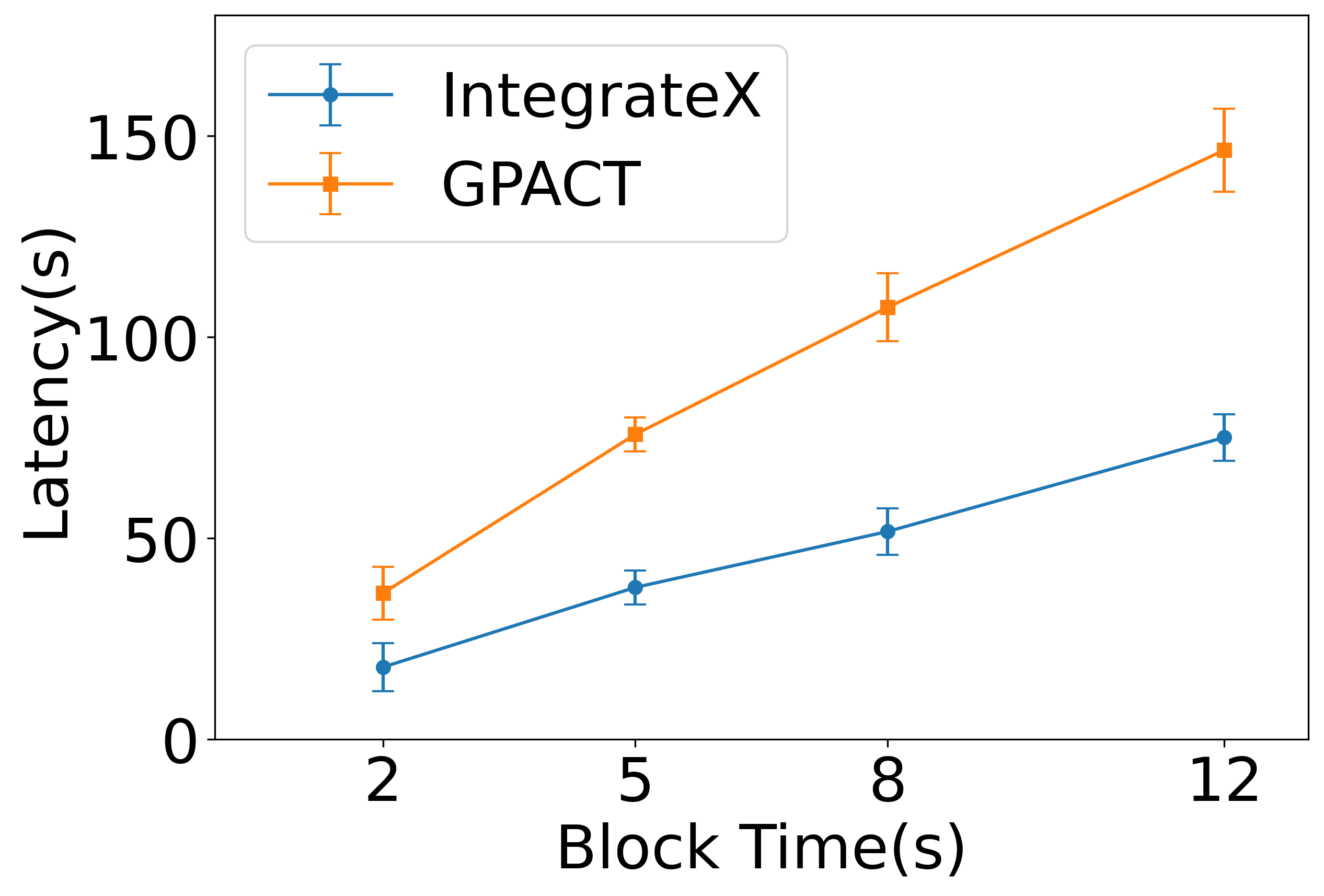}
    \vspace{-6pt}
    \caption{Latency of handling CCSCI in \texttt{IntegrateX} with different block times.}
    \label{diff_bt}
\end{figure} 

\textcolor{red}{
Current EVM-compatible blockchains exhibit a range of block times. The results presented in Figure \ref{diff_bt} show the experiments conducted on the Train-and-Hotel use case under different block time settings. The results indicate that \texttt{IntegrateX} consistently demonstrates significantly lower latency compared to GPACT across all block times. This improvement is primarily due to \texttt{IntegrateX}'s ability to avoid multiple rounds of cross-chain execution during CCSCI, thereby reducing overall latency. As the block time increases, the latency of both \texttt{IntegrateX} and GPACT exhibits an approximately linear growth trend.
}
% The gas consumption on bc1 increases in \texttt{IntegrateX} compared to GPACT, while the gas consumption on bc2 and bc3 decreases. 
% In \texttt{IntegrateX}, the execution chain (bc1) handles all the logic in CCSCI, meaning that the invoked chains (bc2 and bc3) only need to lock and update states. 

% The results show that in this use case, the average latency for GPACT is 76.35 seconds, while the average latency for \texttt{IntegrateX} and \texttt{IntegrateX} without transaction aggregation is 37.25 seconds and 36.68 seconds, respectively. Additionally, \texttt{IntegrateX} consumes less gas compared to GPACT. In the Train-Hotel use case, \texttt{IntegrateX} reduced gas consumption by 1.2\%, while \texttt{IntegrateX} without transaction aggregation (TA) consumed 4.3\% more gas compared to GPACT. Overall, in the train-hotel application, while \texttt{IntegrateX} decreases gas consumption by 1.2\%, it achieves a 51.2\% improvement in efficiency.

\subsubsection{Performance under Different Invocation Complexity}
We evaluate the performance of latency, throughput and gas consumption of \texttt{IntegrateX} and GPACT with different CCSCI complexity.
% We evaluate the performance of \texttt{IntegrateX} without transaction aggregation at the same time.
% Since there are only three chains, multiple cross-chain contracts are invoked when the call tree depth exceeds 2. 
We vary the call tree depth to represent different levels of complexity in cross-chain invocations.
% adjust the complexity of cross-chain invocations and evaluate the gas consumption and latency of \texttt{IntegrateX} and GPACT at different depths. 
% The comparison results are shown in Figure~\ref{depth}.

\begin{figure}[t]
    \centering
    % 第一个子图
    \begin{subfigure}[b]{0.156\textwidth}
        \centering
        \includegraphics[width=\textwidth]{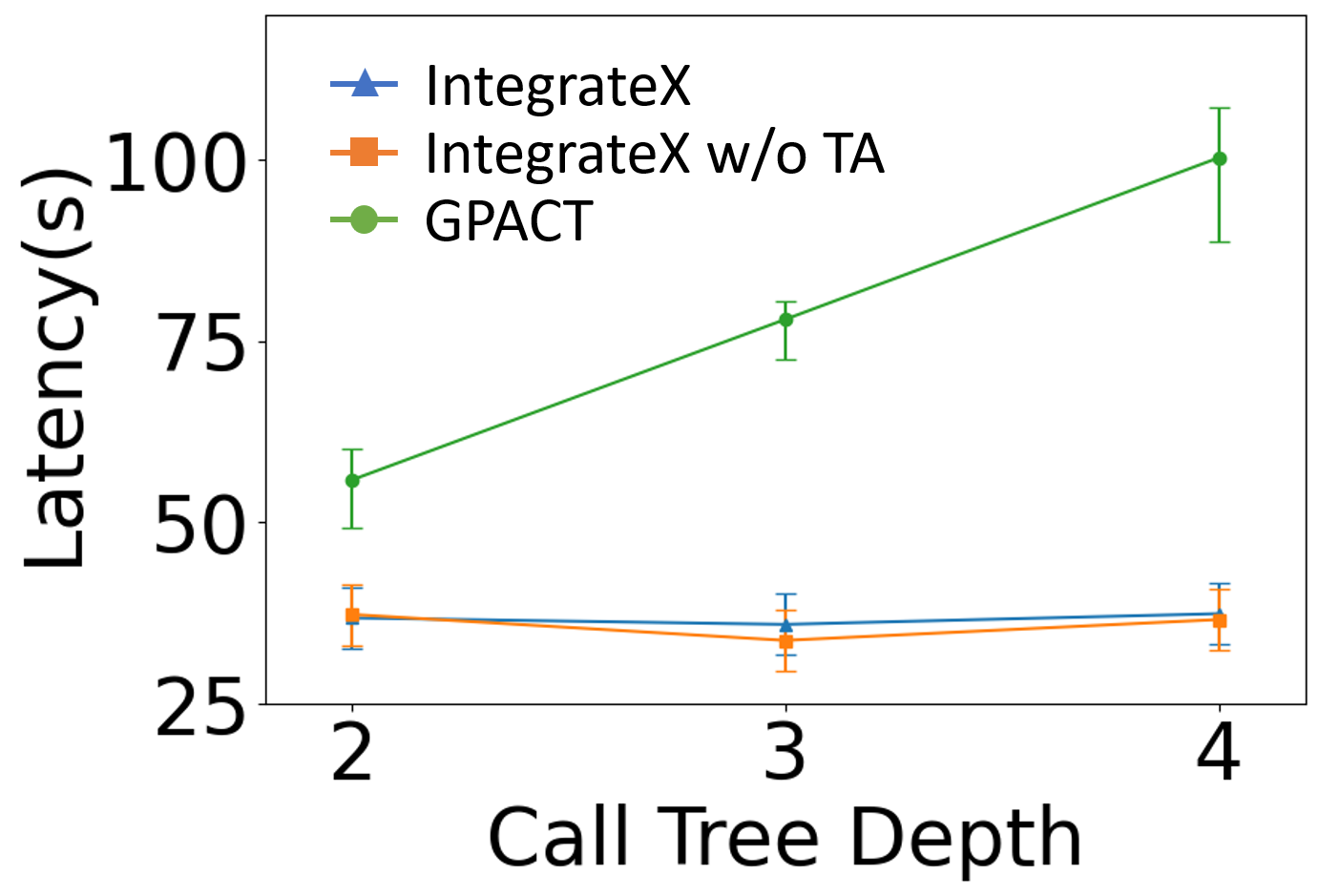}  % 使用你的图片路径
        \caption{Latency}
        \label{dptime}
    \end{subfigure}
    % \hspace{3pt}
    \hfill
    \begin{subfigure}[b]{0.156\textwidth}
        \centering
        \includegraphics[width=\textwidth]{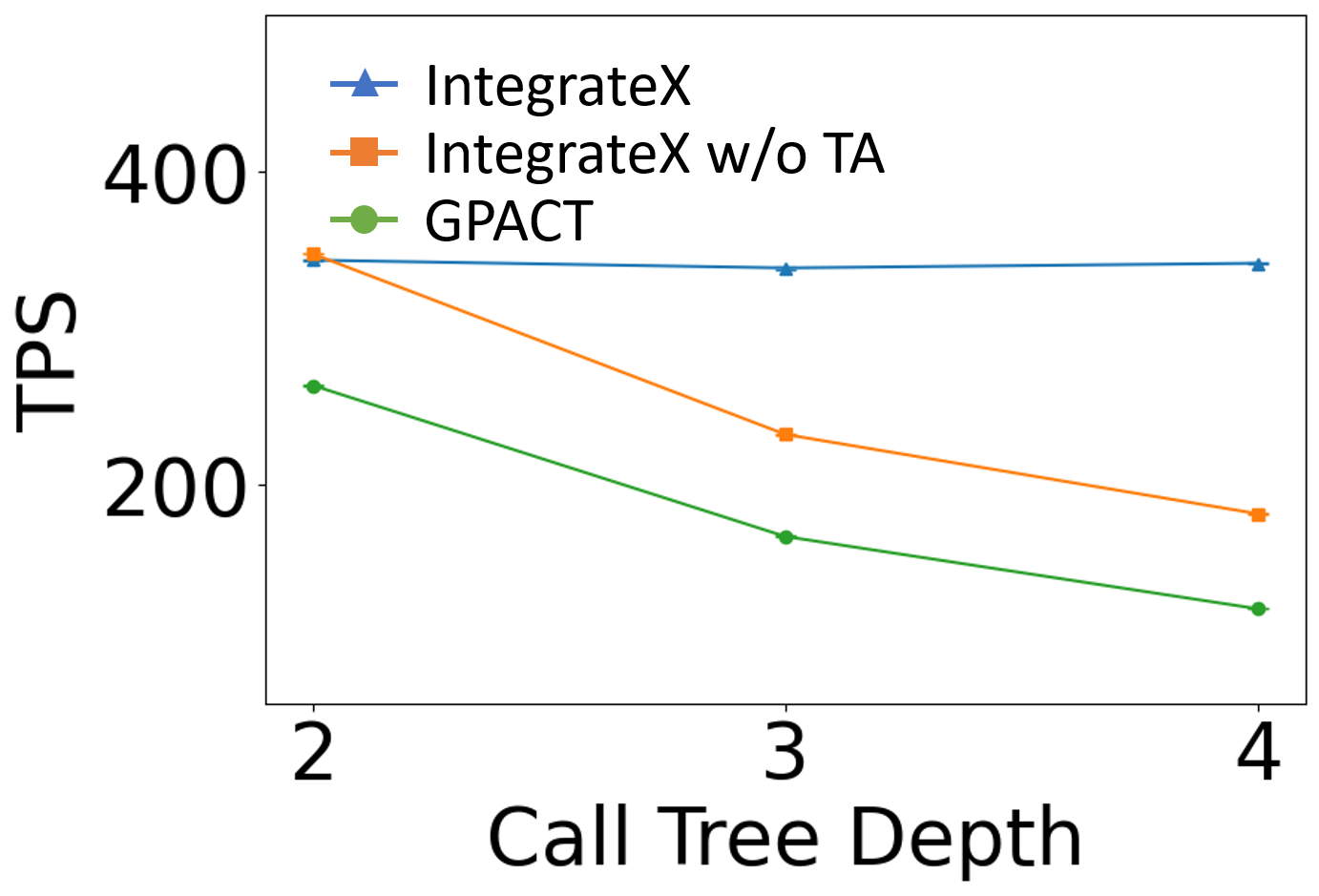}  % 使用你的图片路径
            \caption{Throughput}
        \label{tps}  % 子图标签，用于引用
    \end{subfigure}   
    \hfill
     % 第二个子图
    \begin{subfigure}[b]{0.156\textwidth}
        \centering
        \includegraphics[width=\textwidth]{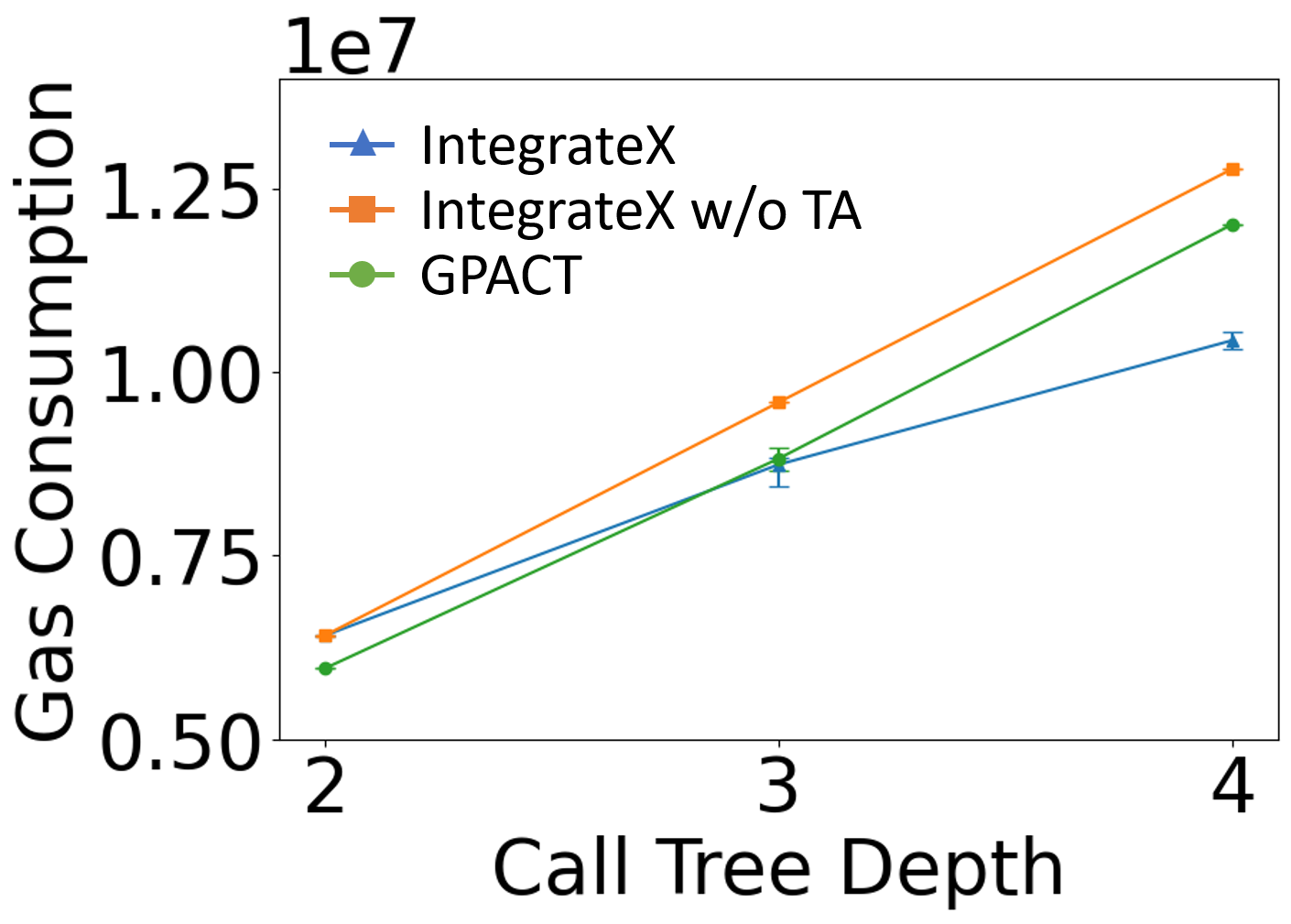}  % 使用你的图片路径
        \caption{Gas consumption}
        \label{dpgas}  % 子图标签，用于引用
    \end{subfigure}   
    \vspace{-6pt}
    \caption{Latency, gas consumption and throughput under different call tree depths.}
    % Gas consumption and latency of IntegrateX and GPACT of different call tree depth.}
    \label{depth}  % 整体图的标签

\end{figure}

The latency results are shown in Figure~\ref{dptime}. Due to the atomic integrated execution mechanism, the latency of \texttt{IntegrateX} and \texttt{IntegrateX} without transaction aggregation (TA) remain stable at around 35 seconds. 
The latency of GPACT, however, increases linearly as the complexity of CCSCI grows.
Specifically, when the call tree depth is 4, \texttt{IntegrateX} significantly reduces latency by 61.2\% compared to GPACT. 
Furthermore, as can be inferred, with invocation complexity further increases, \texttt{IntegrateX} will show greater improvements compared to GPACT. 
% , indicating that with further increases in invocation complexity, \texttt{IntegrateX} will show greater improvements compared to GPACT. 
% When the call tree depth is 2, \texttt{IntegrateX} reduces latency by 33.3\% compared to GPACT. 
% When the call tree depth is 4, \texttt{IntegrateX} reduces latency by 61.2\% compared to GPACT. 

\textcolor{red}{
% In this experiment, each block can contain up to 4096 transactions, and we assume that the block gas limit is not reached.
The throughput results are shown in Figure~\ref{tps}. \texttt{IntegrateX} achieves significantly higher throughput than GPACT across varying call tree depths.
Specifically, when the call tree depth is 2, 3, and 4, \texttt{IntegrateX} outperforms GPACT by 30.4\%, 104.3\%, and 199.1\%, respectively.
This improvement stems from \texttt{IntegrateX}'s transaction aggregation mechanism, which ensures that the number of transactions required to execute CCSCI remains constant, regardless of call tree depth.
As a result, the throughput of \texttt{IntegrateX} remains nearly stable.
In contrast, \texttt{IntegrateX} without transaction aggregation, as well as GPACT, both exhibit a linear increase in the number of required transactions as the call tree depth increases, leading to a corresponding linear decrease in throughput.
Moreover, since integrated execution in \texttt{IntegrateX} requires fewer transactions than GPACT even without aggregation, it consistently achieves better throughput.
Therefore, the performance advantage of \texttt{IntegrateX} becomes increasingly pronounced with deeper call trees.
}

The gas consumption results are shown in Figure~\ref{dpgas}. 
Due to the lower invocation complexity, the gas savings from transaction aggregation are not significant,
when the call tree depth is 3, the gas consumption for CCSCI in \texttt{IntegrateX} is nearly the same compared to GPACT. As the invocation complexity increases, when the call tree depth reaches 4, transaction aggregation reduces gas consumption in \texttt{IntegrateX} by 9.8\% compared to GPACT. And with further increases in invocation complexity, transaction aggregation will reduces more gas consumption in \texttt{IntegrateX}. We also evaluate the gas consumption between \texttt{IntegrateX} and \texttt{IntegrateX} without transaction aggregation. The experimental results show that as the complexity of CCSCI increases, the transaction aggregation mechanism significantly reduces gas consumption in CCSCI. When the call tree depth is 3, the transaction aggregation mechanism reduces gas consumption by 5.4\%, and when the call tree depth is 4, it reduces gas consumption by 19.3\%. This suggests that the transaction aggregation mechanism achieves greater improvements in gas consumption as the complexity of CCSCI increases.

% Overall, as the complexity of cross-chain invocations increases, \texttt{IntegrateX} not only demonstrates significant performance improvements but also exhibits progressively lower gas consumption compared to GPACT.

\subsubsection{Concurrency Performance with Fine-Grained State Lock}
\begin{figure}[t]
    \centering
    \includegraphics[width=0.22\textwidth]{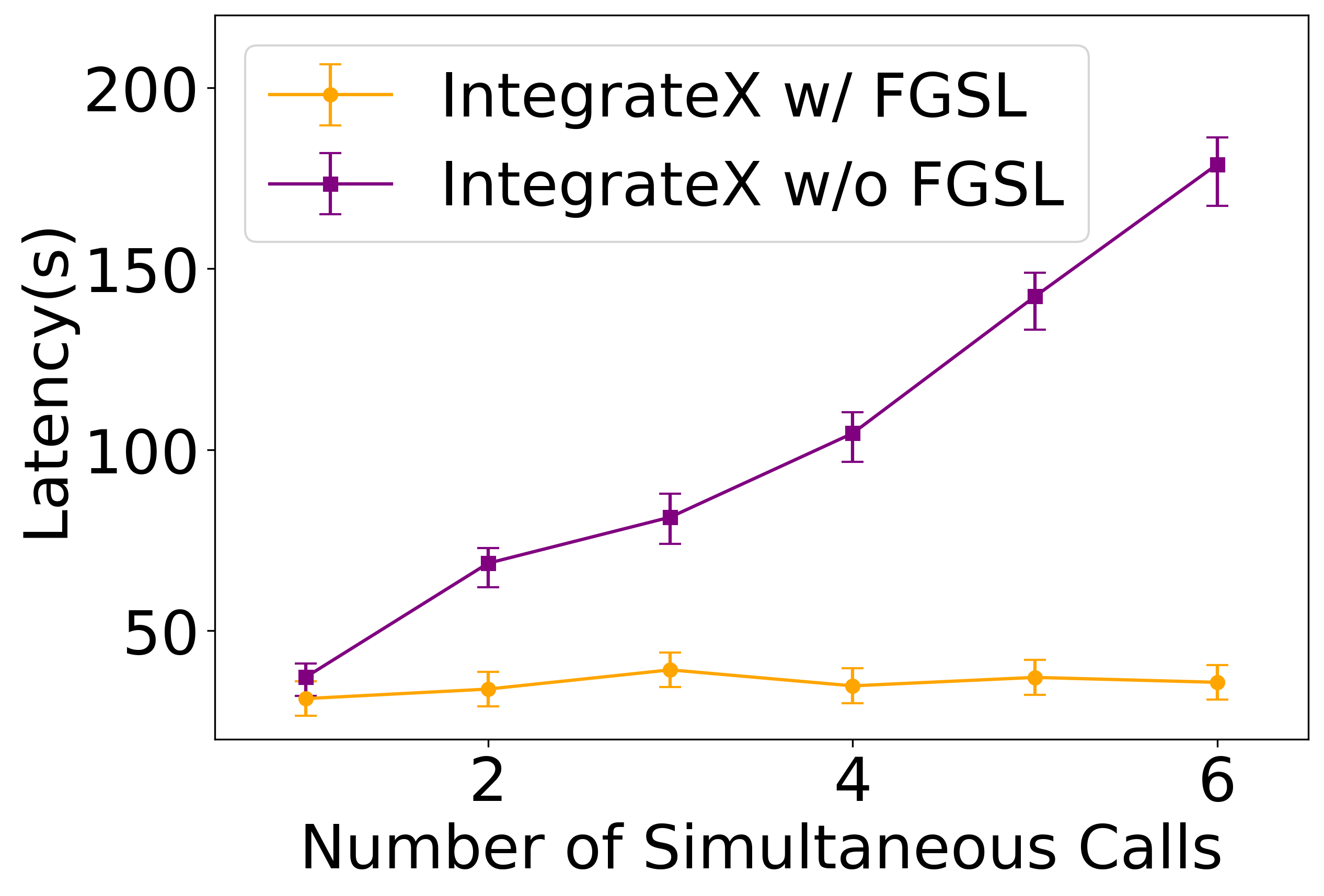}
    \vspace{-6pt}
    \caption{Latency of handling CCSCI in \texttt{IntegrateX} with/without Fine-Grained State Lock (FGSL).}
    \label{fgl_t}
    \vspace{-6pt}
\end{figure}

We now measure the concurrency performance of \texttt{IntegrateX} with and without the fine-grained state lock mechanism. {\textcolor{red} Concurrency refers to the system's ability to execute multiple full cross-chain smart contract invocations in parallel, rather than isolated operations on a single chain. }
% In this experiment, to show the concurrency performance, we compare the latency of \texttt{IntegrateX} with and without the fine-grained state lock mechanism. 
% We ensured that the locked state could successfully handle all the calls, thus avoiding failures due to incomplete calls. 
To show the concurrency performance, we initiate multiple cross-chain calls on the same state simultaneously and compare the time consumed by the cross-chain dApp. 
The experimental results are shown in the Figure~\ref{fgl_t}.

The experimental results show that with fine-grained state lock mechanism, the latency of \texttt{IntegrateX} remains stable at around 35 seconds regardless of the number of simultaneous calls.
This demonstrates that the fine-grained state lock can maintain good transaction concurrency.
% When the number of simultaneous invocations is 2, the average latency of the \texttt{IntegrateX} system without fine-grained state lock is 68.6 seconds, representing an increase of approximately 102.3\%.
On the other side, as the number of simultaneous calls increases, the latency increases linearly in \texttt{IntegrateX} without fine-grained state lock. 
When the number of simultaneous invocations reached 6, the average latency of the \texttt{IntegrateX} system rises to 178.8 seconds, representing an increase of approximately 400.1\%.
% Without fine-grained state lock mechanism, the latency increased linearly with the number of simultaneous invocations. 

\subsubsection{Logic-State Decoupling}
We evaluate the gas consumption during the cross-chain smart contract deployment, using the Train-and-Hotel contracts as an example.
We compare the results of the train and hotel contracts with and without logic-state decoupling.
% logic clone for smart contracts, comparing the logic-state decoupling version with the non-decoupled version. 
% In the train-hotel example, we test two contracts: the train contract, which has more complex logic, and the hotel contract, which stores more states. The experimental results are presented in Table~\ref{lsd}.
The experimental results in Table~\ref{lsd} indicate that, after applying logic-state decoupling, gas consumption during off-chain clone and deployment is reduced by 48.6\% for the train contract and 74.5\% for the hotel contract. 
The main reason is that, the train contract is implemented with more complex logic, while the hotel contract stores more states. 
As can been seen, for contracts with substantial state but relatively simple logic, logic-state decoupling can significantly reduce gas consumption.

\begin{table}[tbh]
\centering
\caption{Gas consumption during cross-chain deployment.}
%\vspace{-10pt}
\begin{tabular}{|c|c|c|}
\hline
\textbf{  Contract Type  }  & \textbf{  w/o LSD 
(Gas)  } & \textbf{  w/ LSD (Gas)  } \\ \hline
Train Contract          & 1,459,626                                & 749,383                              \\ \hline
Hotel Contract          & 1,524,151                                & 389,383                              \\ \hline
\end{tabular}
\label{lsd}
%\vspace{-10pt}
\end{table}

\subsubsection{On-chain Operation During Off-chain Clone and Deployment}
\begin{figure}[t]
    \centering
    % 第一个子图
    \begin{subfigure}[b]{0.22\textwidth}
        \centering
        \includegraphics[width=\textwidth]{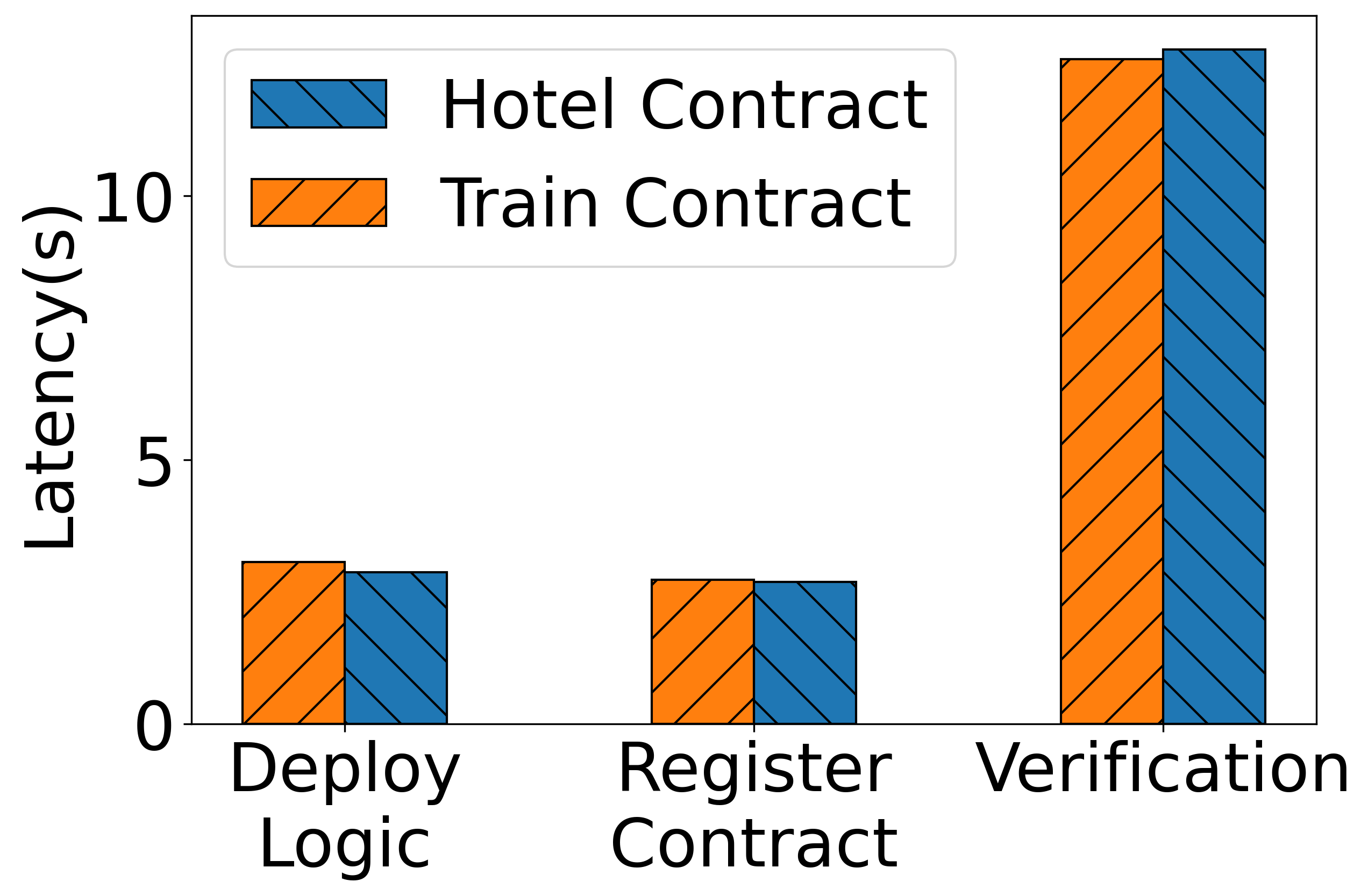}  % 使用你的图片路径
        \caption{Latency}
        \label{ocoptime}
    \end{subfigure}
    % \hspace{3pt}
    \hfill
     % 第二个子图
    \begin{subfigure}[b]{0.22\textwidth}
        \centering
        \includegraphics[width=\textwidth]{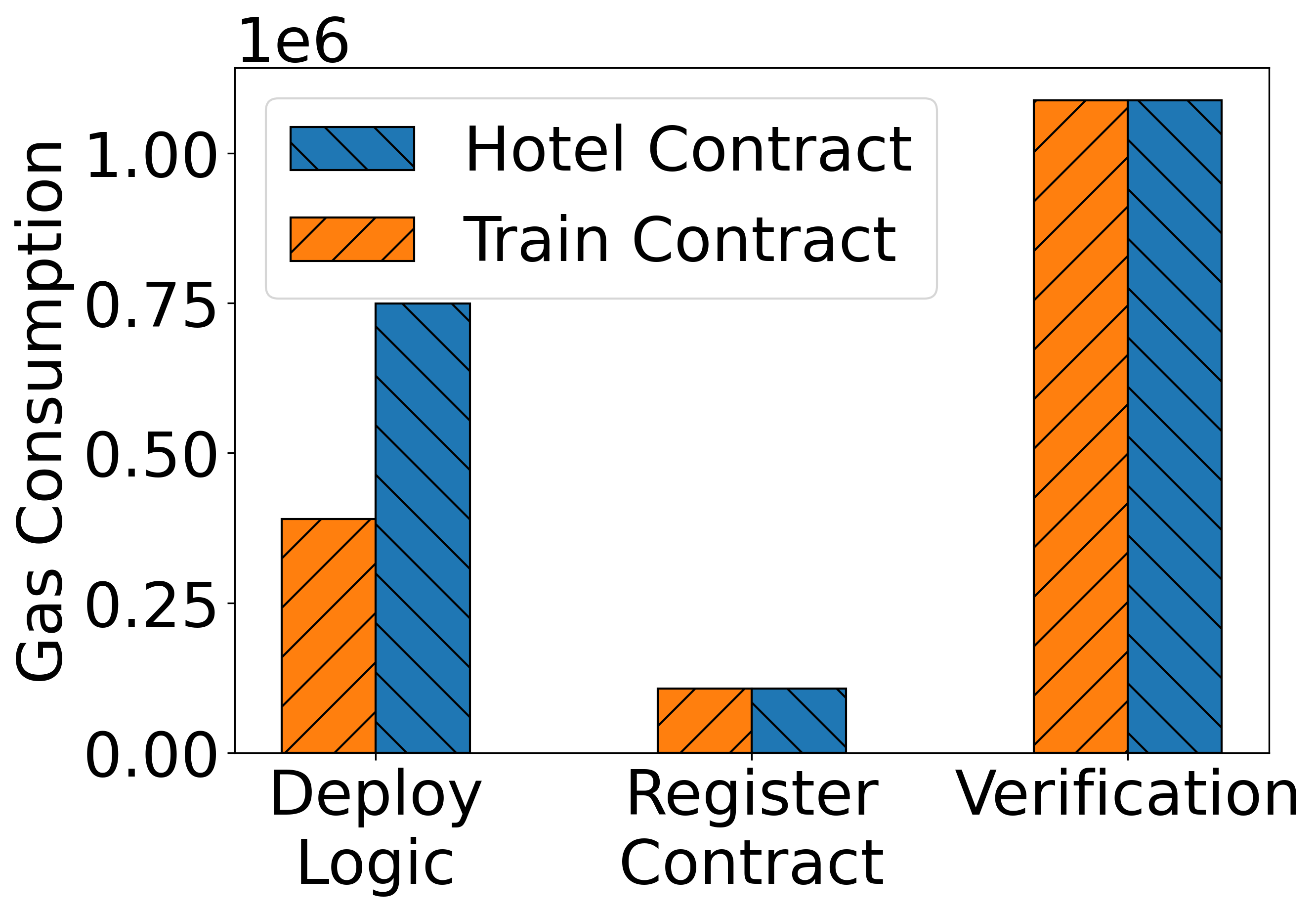}  % 使用你的图片路径
        \caption{Gas consumption}
        \label{ocopgas}  % 子图标签，用于引用
    \end{subfigure}
    \vspace{-6pt}
    \caption{Latency and gas consumption of different on-chain operations during off-chain clone and deployment.}
    % TA: transaction aggregation.}
    \label{ocop}  % 整体图的标签
    \vspace{-6pt}
\end{figure}

\textcolor{red}{
We evaluate the gas consumption and latency associated with the on-chain operations during the off-chain clone and deployment phase of the Hybrid Cross-Chain Smart Contract Deployment Protocol. The results are shown in Figure~\ref{ocop}.
Figure~\ref{ocoptime} presents the latency of different operations, which are nearly identical for both the \emph{Hotel} and \emph{Train} contracts. 
Because both the Deploy Logic and Register Contract operations complete within a single on-chain transaction, their latency remains low. Verification, by contrast, requires cross-chain validation and therefore incurs higher latency.
% Since the logic deployment and contract registration can be completed within a single block, their average latency is typically short.
% % approximately half a block time.
% In contrast, the on-chain verification requires at least three blocks to complete, resulting in a longer average latency.
% of roughly two and a half block times.
Figure~\ref{ocopgas} shows the gas consumption of these operations. The \emph{Train} contract consumes more gas for logic deployment due to its more complex functionality. However, for contract registration and verification, both contracts perform similar operations, leading to nearly identical gas usage.
It is important to note that these operations are one-time costs. 
Overall, the results demonstrate that the on-chain overhead is reasonably lightweight and practical in terms of both gas and latency.
}

\section{Discussions and Limitations}\label{disscussion}

\noindent
\textbf{Potential Application Scenarios.} 
Beyond the example of the train-and-hotel problem mentioned in this paper, \texttt{IntegrateX} can be extended to a wide range of application scenarios. 
One promising use case is cross-chain flash loans \cite{tefagh2021ccfl}. 
Flash loans are atomic, uncollateralized lending protocols that allow users to borrow funds at nearly zero cost, perform other operations, and then repay the loan. 
However, these processes require the guarantee of overall atomicity, meaning that either all steps succeed or they all fail. 
Due to this requirement for atomicity, existing flash loan protocols are limited to intra-chain operations. 
With \texttt{IntegrateX}, which provides overall atomicity for cross-chain dApps, users can efficiently perform cross-chain flash loan operations.
Other promising application scenarios include, but are not limited to, cross-chain atomic arbitrage and cross-chain supply chain management.

\vspace{3pt}
\noindent
\textbf{Learning Cost for Developers.} 
The logic-state decoupling and fine-grained state lock mechanisms may slightly increase the development learning curve for smart contract developers. Fortunately, we have proposed a set of guidelines to assist developers, and the mechanisms are flexible (as discussed in Section \ref{subsec:LSD} and \ref{subsec:lock}), allowing developers to freely decide whether to implement them. Furthermore, we can provide formal documentation, SDKs, and other resources (following existing standards such as Wormhole \cite{wormhole} and IBC \cite{cosmos2019}) to guide developers in secure and efficient development and auditing, thereby reducing the learning curve. Additionally, incentive mechanisms (e.g., token rewards) are widely adopted in the industry to encourage developers to build and utilize our system. In future research, we could even explore AI-based semi-automated smart contract tools to further address this challenge.

Moreover, logic-state decoupling offers several additional benefits. For instance, it facilitates modular programming principles in smart contract development, which helps reduce subsequent upgrade and maintenance costs while improving contract security. When an issue arises in a specific contract module, developers can conduct targeted audits and resolve vulnerabilities efficiently.

\vspace{3pt}
\noindent
\textbf{Support for Heterogeneous Chains.} 
\texttt{IntegrateX} currently supports blockchains that run different consensus protocols but share the same smart contract execution environment. 
As mentioned in the paper, to ensure cross-chain transaction security across blockchains with different consensus protocols, \texttt{IntegrateX} waits until consensus on the source chain is finalized (or highly likely to be finalized) before committing the cross-chain transaction to the target chain. 
Additionally, while this paper focuses on \texttt{IntegrateX}’s implementation on EVM-compatible blockchains, it can also be modified to operate between non-EVM-compatible blockchains that share the same smart contract execution environment.

However, a current limitation of \texttt{IntegrateX} is that it cannot operate between blockchains with different smart contract execution environments. 
Fortunately, this issue could potentially be addressed using advanced techniques such as code virtualization \cite{virtualization}. 
Expanding \texttt{IntegrateX} to support integrated execution across blockchains with different smart contract environments is a future research direction we aim to explore.

\vspace{3pt}
\noindent
\textbf{Trade-Off Between Flexibility and Load Balancing.} 
In \texttt{IntegrateX}, cross-chain dApp providers have the flexibility to select any chain as the execution chain for integrated execution, based on their preferences. 
However, this flexibility may introduce a potential issue: if many cross-chain dApp providers choose the same chain as the execution chain, that chain could become a hotspot, potentially degrading its performance.
One possible solution is for a third party (e.g., \texttt{IntegrateX}) to manage load balancing by selecting the execution chain on behalf of the cross-chain dApp providers. 
However, this approach could introduce centralization risks. 
Additionally, as discussed in the paper, developers' choice of which chain to run their dApps on often involves considerations beyond performance, such as ecosystem compatibility and business partnerships.
How to better achieve load balancing and how to strike a trade-off between performance and flexibility are important questions that warrant future research.

\vspace{3pt}
\noindent
\textbf{Mitigating Malicious Application Layer Components.}
In public blockchain scenarios, there are common strategies to mitigate malicious behavior from application layer components (e.g., dApp providers, users). 
For instance, a malicious cross-chain dApp provider might attempt to maliciously lock certain states to prevent their usage by others. 
Such behavior can be countered using contract-based authorization or blacklisting mechanisms (widely used in existing dApp development \cite{etherscan}). 
For example, an intra-chain dApp provider can pre-arrange with a cross-chain dApp provider and authorize trusted cross-chain dApp providers (through their associated addresses) in their contracts, allowing only authorized cross-chain dApp providers to invoke and lock their states. 
Similarly, intra-chain dApp providers can blacklist specific cross-chain dApp providers in their contracts to block their interactions.
Additionally, gas fee mechanisms can serve as a deterrent to malicious application layer components attempting to launch flooding attacks against the blockchain.

\vspace{3pt}
\noindent
\textbf{Cross-Chain vs. Cross-Shard.}
Some existing works have explored the issue of cross-shard smart contract handling \cite{qi2024lightcross, li2022jenga}. 
However, cross-chain and cross-shard scenarios are fundamentally different, and their solutions cannot be directly applied to cross-chain contexts. 
The primary reasons are as follows: First, research on blockchain sharding typically involves modifications to the underlying system \cite{cochain,spchain}. 
In contrast, a key requirement of cross-chain protocols or systems is that they must not require modifications to the underlying blockchains, ensuring better compatibility with existing blockchain systems. 
Second, a blockchain sharding system usually has a beacon chain~\cite{lbchain,spiralshard} responsible for coordinating progress across shards, which is impractical in cross-chain protocols. 
After all, in cross-chain scenarios, each blockchain essentially belongs to a different system. 
Finally, in cross-chain contexts, blockchains are often heterogeneous (e.g., different consensus protocols), which is uncommon in blockchain sharding systems.

\vspace{3pt}
\noindent
\textbf{Inter-Chain Shared Security.}
In \texttt{IntegrateX}, each blockchain is assumed to have a proportion of malicious nodes lower than its fault tolerance threshold (i.e., they are secure). 
This is a widely accepted assumption in most existing works. 
However, recent research has begun exploring how to achieve secure cross-chain interoperability protocols in scenarios where individual blockchains may not be secure, by sharing security across multiple blockchains \cite{sheng2023trustboost}.
\texttt{IntegrateX} could adopt similar ideas through modifications to achieve shared security among blockchains. 
However, how to design and implement inter-chain shared security within \texttt{IntegrateX} while still maintaining efficiency is an important direction for future research.

\section{Conclusion}\label{conclusion}

% In this paper, we propose \texttt{IntegrateX}, an interoperability system that supports efficient cross-chain smart contract invocations with overall atomicity. 
% \texttt{IntegrateX} comprises two protocols and is designed to facilitate efficient complex cross-chain dApps while ensuring atomicity. 
% We introduced an innovative design and conducted comprehensive experiments, with the results indicating that \texttt{IntegrateX} can enhance call efficiency while maintaining atomicity. Moreover, the more complex the cross-chain calls, the more significant the efficiency gains brought by \texttt{IntegrateX}. In certain complex call scenarios, \texttt{IntegrateX} not only improves efficiency but also reduces gas consumption. 
% % In future research, we will focus on enhancing the system's resilience against attacks, as well as further optimizing the efficiency of \texttt{IntegrateX}.

In this paper, we introduce \texttt{IntegrateX}, a system that enhances cross-chain interoperability by ensuring overall atomicity and efficiency in cross-chain smart contract invocations across EVM-compatible blockchains. 
The proposed hybrid cross-chain smart contract deployment protocol enables the logic contract to be cloned and deployed onto a single blockchain efficiently and securely.
The proposed cross-chain smart contract integrated execution protocol allows efficient intra-chain integrated execution for cross-chain smart contract invocations while guarantee overall atomicity. 
% By integrating the execution logic of cross-chain applications onto a single blockchain, IntegrateX reduces the need for multiple rounds of cross-chain execution, addressing inefficiencies related to latency and gas consumption. 
% Our protocols—Hybrid Cross-Chain Smart Contract Deployment and Cross-Chain Smart Contract Integrated Execution—improve performance, concurrency, and scalability. 
Experimental results show that, compared to the state-of-the-art baseline, \texttt{IntegrateX} is able to reduce a significant portion of latency while maintaining low gas cost and high concurrency, particularly excelling in complex cross-chain interactions. 

% \clearpage
\bibliographystyle{IEEEtran}
\bibliography{sample}

\begin{thebibliography}{10}
\providecommand{\url}[1]{#1}
\csname url@samestyle\endcsname
\providecommand{\newblock}{\relax}
\providecommand{\bibinfo}[2]{#2}
\providecommand{\BIBentrySTDinterwordspacing}{\spaceskip=0pt\relax}
\providecommand{\BIBentryALTinterwordstretchfactor}{4}
\providecommand{\BIBentryALTinterwordspacing}{\spaceskip=\fontdimen2\font plus
\BIBentryALTinterwordstretchfactor\fontdimen3\font minus \fontdimen4\font\relax}
\providecommand{\BIBforeignlanguage}[2]{{%
\expandafter\ifx\csname l@#1\endcsname\relax
\typeout{** WARNING: IEEEtran.bst: No hyphenation pattern has been}%
\typeout{** loaded for the language `#1'. Using the pattern for}%
\typeout{** the default language instead.}%
\else
\language=\csname l@#1\endcsname
\fi
#2}}
\providecommand{\BIBdecl}{\relax}
\BIBdecl

\bibitem{bitcoin}
S.~Nakamoto, ``Bitcoin: A peer-to-peer electronic cash system,'' \emph{Satoshi Nakamoto}, 2008.

\bibitem{eth}
V.~Buterin \emph{et~al.}, ``Ethereum white paper,'' \emph{GitHub repository}, vol.~1, pp. 22--23, 2013.

\bibitem{belchior2021pastSV}
R.~Belchior, A.~Vasconcelos, S.~Guerreiro, and M.~Correia, ``A survey on blockchain interoperability: Past, present, and future trends,'' \emph{ACM Comput. Surv.}, pp. 1--41, 2021.

\bibitem{huang2021survey}
H.~Huang, W.~Kong, S.~Zhou, Z.~Zheng, and S.~Guo, ``A survey of state-of-the-art on blockchains: Theories, modelings, and tools,'' \emph{{ACM} Comput. Surv.}, vol.~54, no.~2, pp. 44:1--44:42, 2021.

\bibitem{wenkai2022defiSV}
W.~Li, J.~Bu, X.~Li, H.~Peng, Y.~Niu, and Y.~Zhang, ``A survey of defi security: Challenges and opportunities,'' \emph{Journal of King Saud University - Computer and Information Sciences}, vol.~34, no. 10, Part B, pp. 10\,378--10\,404, 2022.

\bibitem{nadini2021mapping}
M.~Nadini, L.~Alessandretti, F.~Di~Giacinto, M.~Martino, L.~M. Aiello, and A.~Baronchelli, ``Mapping the nft revolution: market trends, trade networks, and visual features,'' \emph{Scientific reports}, vol.~11, no.~1, p. 20902, 2021.

\bibitem{ethereum_evm}
Ethereum, ``Ethereum virtual machine (evm) documentation,'' \url{https://ethereum.org/en/developers/docs/evm/}, 2024, accessed: Feb. 2025.

\bibitem{chaintvl}
DefiLlama, ``Total value locked all chains,'' \url{https://defillama.com/chains}, 2024, accessed: Feb. 2025.

\bibitem{OU2022OV}
W.~Ou, S.~Huang, J.~Zheng, Q.~Zhang, G.~Zeng, and W.~Han, ``An overview on cross-chain: Mechanism, platforms, challenges and advances,'' \emph{Computer Networks}, vol. 218, p. 109378, 2022.

\bibitem{Falazi2024crosschain}
G.~Falazi, U.~Breitenb\"{u}cher, F.~Leymann, and S.~Schulte, ``Cross-chain smart contract invocations: A systematic multi-vocal literature review,'' \emph{ACM Comput. Surv.}, vol.~56, no.~6, pp. 1--38, 2024.

\bibitem{train}
V.~Buterin, ``What is the train-and-hotel problem?'' \url{https://vitalik.eth.limo/general/2017/12/31/sharding_faq.html\#what-is-the-train-and-hotel-problem}, 2017, accessed: Feb. 2025.

\bibitem{lampson1993twopc}
B.~Lampson and D.~Lomet, ``A new presumed commit optimization for two phase commit,'' in \emph{19th International Conference on Very Large Data Bases (VLDB'93)}, 1993, pp. 630--640.

\bibitem{nissl2021towards}
M.~Nissl, E.~Sallinger, S.~Schulte, and M.~Borkowski, ``Towards cross-blockchain smart contracts,'' in \emph{2021 IEEE International Conference on Decentralized Applications and Infrastructures (DAPPS)}.\hskip 1em plus 0.5em minus 0.4em\relax IEEE, 2021, pp. 85--94.

\bibitem{wood2016polkadot}
G.~Wood, ``Polkadot: Vision for a heterogeneous multi-chain framework,'' \emph{White paper}, vol.~21, no. 2327, pp. 2327--4662, 2016.

\bibitem{cosmos2019}
J.~Kwon and E.~Buchman, ``Cosmos white paper,'' White Paper, 2019.

\bibitem{abebe2019enabling}
E.~Abebe, D.~Behl, C.~Govindarajan, Y.~Hu, D.~Karunamoorthy, P.~Novotny, V.~Pandit, V.~Ramakrishna, and C.~Vecchiola, ``Enabling enterprise blockchain interoperability with trusted data transfer (industry track),'' in \emph{Proceedings of the 20th International Middleware Conference Industrial Track}, 2019, p. 29–35.

\bibitem{darshan2023an}
M.~Darshan, M.~Amet, G.~Srivastava, and J.~Crichigno, ``An architecture that enables cross-chain interoperability for next-gen blockchain systems,'' \emph{IEEE Internet of Things Journal}, vol.~10, no.~20, pp. 18\,282--18\,291, 2023.

\bibitem{reigsbergen2023demo}
D.~Reijsbergen, A.~Maw, J.~Zhang, T.~T.~A. Dinh, and A.~Datta, ``Demo: Piechain - a practical blockchain interoperability framework,'' in \emph{2023 IEEE 43rd International Conference on Distributed Computing Systems (ICDCS)}, 2023, pp. 1021--1024.

\bibitem{ghosh2021leveraging}
B.~C. Ghosh, T.~Bhartia, S.~K. Addya, and S.~Chakraborty, ``Leveraging public-private blockchain interoperability for closed consortium interfacing,'' in \emph{IEEE INFOCOM 2021 - IEEE Conference on Computer Communications}, 2021, pp. 1--10.

\bibitem{garoffolo2020zendoo}
A.~Garoffolo, D.~Kaidalov, and R.~Oliynykov, ``Zendoo: a zk-snark verifiable cross-chain transfer protocol enabling decoupled and decentralized sidechains,'' in \emph{2020 IEEE 40th International Conference on Distributed Computing Systems (ICDCS)}, 2020, pp. 1257--1262.

\bibitem{robinson2021general}
P.~Robinson and R.~Ramesh, ``General purpose atomic crosschain transactions,'' in \emph{2021 3rd Conference on blockchain research \& applications for innovative networks and services (BRAINS)}.\hskip 1em plus 0.5em minus 0.4em\relax IEEE, 2021, pp. 61--68.

\bibitem{atomic-ibc}
AtomicIBC, ``Atomic ibc,'' \url{https://informal.systems/blog/atomic-ibc}, 2024, accessed: Feb. 2025.

\bibitem{chen2024atomci}
Y.~Chen, A.~Asheralieva, and X.~Wei, ``Atomci: A new system for the atomic cross-chain smart contract invocation spanning heterogeneous blockchains,'' \emph{IEEE Transactions on Network Science and Engineering}, vol.~11, no.~3, pp. 2782--2796, 2024.

\bibitem{INTEGRATEX}
``Implementation of integratex,'' \url{https://github.com/B-Above/INtegrateX}, 2024.

\bibitem{2019atomicBEswap}
L.~Lys, A.~Micoulet, and M.~Potop-Butucaru, ``Atomic swapping bitcoins and ethers,'' in \emph{2019 38th Symposium on Reliable Distributed Systems (SRDS)}, 2019, pp. 372--3722.

\bibitem{Xu2021htlc}
J.~Xu, D.~Ackerer, and A.~Dubovitskaya, ``A game-theoretic analysis of cross-chain atomic swaps with htlcs,'' in \emph{2021 IEEE 41st International Conference on Distributed Computing Systems (ICDCS)}, 2021, pp. 584--594.

\bibitem{Luo2024crosschannel}
X.~Luo, K.~Xue, Q.~Sun, and J.~Lu, ``Crosschannel: Efficient and scalable cross-chain transactions through cross-and-off-blockchain micropayment channel,'' \emph{IEEE Transactions on Dependable and Secure Computing}, pp. 1--15, 2024.

\bibitem{Manevich2022ccas}
Y.~Manevich and A.~Akavia, ``Cross chain atomic swaps in the absence of time via attribute verifiable timed commitments,'' in \emph{2022 IEEE 7th European Symposium on Security and Privacy (EuroS\&P)}, 2022, pp. 606--625.

\bibitem{tian2021enabling}
H.~Tian, K.~Xue, X.~Luo, S.~Li, J.~Xu, J.~Liu, J.~Zhao, and D.~S. Wei, ``Enabling cross-chain transactions: A decentralized cryptocurrency exchange protocol,'' \emph{IEEE Transactions on Information Forensics and Security}, vol.~16, pp. 3928--3941, 2021.

\bibitem{herlihy2018atomic}
M.~Herlihy, ``Atomic cross-chain swaps,'' in \emph{Proceedings of the 2018 ACM symposium on principles of distributed computing}, 2018, pp. 245--254.

\bibitem{deshpande2020privacy}
A.~Deshpande and M.~Herlihy, ``Privacy-preserving cross-chain atomic swaps,'' in \emph{International Conference on Financial Cryptography and Data Security}.\hskip 1em plus 0.5em minus 0.4em\relax Springer, 2020, pp. 540--549.

\bibitem{thyagarajan2022universal}
S.~A. Thyagarajan, G.~Malavolta, and P.~Moreno-Sanchez, ``Universal atomic swaps: Secure exchange of coins across all blockchains,'' in \emph{2022 IEEE Symposium on Security and Privacy (SP)}.\hskip 1em plus 0.5em minus 0.4em\relax IEEE, 2022, pp. 1299--1316.

\bibitem{chen2024pacdam}
Y.~Chen, J.-N. Liu, A.~Yang, J.~Weng, M.-R. Chen, Z.~Liu, and M.~Li, ``Pacdam: Privacy-preserving and adaptive cross-chain digital asset marketplace,'' \emph{IEEE Internet of Things Journal}, vol.~11, no.~8, pp. 13\,424--13\,436, 2024.

\bibitem{yin2022sidechain}
L.~Yin, J.~Xu, and Q.~Tang, ``Sidechains with fast cross-chain transfers,'' \emph{IEEE Transactions on Dependable and Secure Computing}, vol.~19, no.~6, pp. 3925--3940, 2022.

\bibitem{zamyatin2019xclaim}
A.~Zamyatin, D.~Harz, J.~Lind, P.~Panayiotou, A.~Gervais, and W.~Knottenbelt, ``Xclaim: Trustless, interoperable, cryptocurrency-backed assets,'' in \emph{2019 IEEE Symposium on Security and Privacy (SP)}, 2019, pp. 193--210.

\bibitem{liu2021hyperservice}
Z.~Liu, Y.~Xiang, J.~Shi, P.~Gao, H.~Wang, X.~Xiao, B.~Wen, Q.~Li, and Y.-C. Hu, ``Make web3. 0 connected,'' \emph{IEEE transactions on dependable and secure computing}, vol.~19, no.~5, pp. 2965--2981, 2021.

\bibitem{tao2023atomicity}
Y.~Tao, B.~Li, and B.~Li, ``On atomicity and confidentiality across blockchains under failures,'' \emph{IEEE Transactions on Knowledge and Data Engineering}, pp. 766--780, 2024.

\bibitem{weterkamp2023instant}
M.~Westerkamp and A.~Küpper, ``Instant function calls using synchronized cross-blockchain smart contracts,'' \emph{IEEE Transactions on Network and Service Management}, vol.~20, no.~3, pp. 2136--2150, 2023.

\bibitem{hu2024ivyredaction}
S.~Hu, M.~Li, J.~Weng, J.-N. Liu, J.~Weng, and Z.~Li, ``Ivyredaction: Enabling atomic, consistent and accountable cross-chain rewriting,'' \emph{IEEE Transactions on Dependable and Secure Computing}, vol.~21, no.~4, pp. 3883--3900, 2024.

\bibitem{Multi-Chain-Atomic-Commits}
``Heterogeneous paxos and multi-chain atomic commits,'' \url{https://anoma.net/blog/heterogeneous-paxos-and-multi-chain-atomic-commits}, 2024, accessed: Feb. 2025.

\bibitem{westerkamp2019verifiable}
M.~Westerkamp, ``Verifiable smart contract portability,'' in \emph{2019 IEEE International Conference on Blockchain and Cryptocurrency (ICBC)}, 2019, pp. 1--9.

\bibitem{fynn2020smom}
E.~Fynn, A.~Bessani, and F.~Pedone, ``Smart contracts on the move,'' in \emph{2020 50th Annual IEEE/IFIP International Conference on Dependable Systems and Networks (DSN)}, 2020, pp. 233--244.

\bibitem{westerkamp2022smartsync}
M.~Westerkamp and A.~Küpper, ``Smartsync: Cross-blockchain smart contract interaction and synchronization,'' in \emph{2022 IEEE International Conference on Blockchain and Cryptocurrency (ICBC)}, 2022, pp. 1--9.

\bibitem{bsc_whitepaper}
B.~S. Chain, ``Binance smart chain whitepaper,'' \url{https://github.com/bnb-chain/whitepaper/blob/master/WHITEPAPER.md}, accessed: Feb. 2025.

\bibitem{sheng2023trustboost}
P.~Sheng, X.~Wang, S.~Kannan, K.~Nayak, and P.~Viswanath, ``Trustboost: Boosting trust among interoperable blockchains,'' in \emph{Proceedings of the 2023 ACM SIGSAC Conference on Computer and Communications Security}, 2023, pp. 1571--1584.

\bibitem{tas2023interchain}
E.~N. Tas, R.~Han, D.~Tse, and M.~Yu, ``Interchain timestamping for mesh security,'' in \emph{Proceedings of the 2023 ACM SIGSAC Conference on Computer and Communications Security}, 2023, pp. 1585--1599.

\bibitem{pbft}
M.~Castro and B.~Liskov, ``Practical byzantine fault tolerance,'' in \emph{Proceedings of the 3rd {USENIX} Symposium on Operating Systems Design and Implementation (OSDI 99)}, 1999.

\bibitem{li2022jenga}
M.~Li, Y.~Lin, J.~Zhang, and W.~Wang, ``Jenga: Orchestrating smart contracts in sharding-based blockchain for efficient processing,'' in \emph{Proceedings of the 42nd International Conference on Distributed Computing Systems (ICDCS 22)}, 2022.

\bibitem{feist2019slither}
J.~Feist, G.~Grieco, and A.~Groce, ``Slither: A static analysis framework for smart contracts,'' in \emph{2019 IEEE/ACM 2nd International Workshop on Emerging Trends in Software Engineering for Blockchain (WETSEB)}, 2019, pp. 8--15.

\bibitem{merkle1987}
R.~C. Merkle, ``A digital signature based on a conventional encryption function,'' in \emph{Conference on the Theory and Application of Crypto- graphic Techniques.}, 1987, pp. 369--378.

\bibitem{binance_whitepaper}
B.~Chain, ``{Binance Chain Whitepaper},'' \url{https://github.com/bnb-chain/whitepaper/blob/master/WHITEPAPER.md}, accessed: Jun. 18, 2025.

\bibitem{ye2020bitxhub}
S.~Ye, X.~Wang, C.~Xu \emph{et~al.}, ``Bitxhub: a heterogeneous blockchain interoperability platform based on sidechain relaying [j],'' \emph{Computer Science}, vol.~47, no.~6, pp. 294--302, 2020.

\bibitem{keil2000polygon}
J.~M. Keil, ``Polygon decomposition.'' \emph{Handbook of computational geometry}, vol.~2, pp. 491--518, 2000.

\bibitem{carver2010optimism}
C.~S. Carver, M.~F. Scheier, and S.~C. Segerstrom, ``Optimism,'' \emph{Clinical psychology review}, vol.~30, no.~7, pp. 879--889, 2010.

\bibitem{tefagh2021ccfl}
M.~Tefagh, F.~Bagheri, A.~Khajehpour, and M.~Abdi, ``Atomic bonded cross-chain debt,'' in \emph{Proceedings of the 2020 3rd International Conference on Blockchain Technology and Applications}, ser. ICBTA '20.\hskip 1em plus 0.5em minus 0.4em\relax New York, NY, USA: Association for Computing Machinery, 2021, p. 50–54.

\bibitem{wormhole}
Wormhole, ``Wormhole docs,'' \url{https://wormhole.com/docs/}, 2024, accessed: Feb. 2025.

\bibitem{virtualization}
ChainsAtlas, ``Chainsatlas virtualizatiom unit,'' \url{https://www.chainsatlas.com/}, 2024, accessed: Feb. 2025.

\bibitem{etherscan}
Ethereum, ``Etherscan,'' \url{https://etherscan.io}, 2024, accessed: Feb. 2025.

\bibitem{qi2024lightcross}
X.~Qi and Y.~Li, ``Lightcross: Sharding with lightweight cross-shard execution for smart contracts,'' in \emph{IEEE INFOCOM 2024-IEEE Conference on Computer Communications}.\hskip 1em plus 0.5em minus 0.4em\relax IEEE, 2024, pp. 1681--1690.

\bibitem{cochain}
M.~Li, Y.~Lin, J.~Zhang, and W.~Wang, ``Cochain: High concurrency blockchain sharding via consensus on consensus,'' in \emph{Proceedings of the 42nd {IEEE} Conference on Computer Communications (INFOCOM 23)}, 2023.

\bibitem{spchain}
M.~Li, Y.~Lin, W.~Wang, and J.~Zhang, ``Sp-chain: Boosting intrashard and cross-shard security and performance in blockchain sharding,'' \emph{IEEE Internet of Things Journal}, vol.~12, no.~15, pp. 31\,737--31\,753, 2025.

\bibitem{lbchain}
M.~Li, W.~Wang, and J.~Zhang, ``Lb-chain: Load-balanced and low-latency blockchain sharding via account migration,'' \emph{IEEE Transactions on Parallel and Distributed Systems}, vol.~34, no.~10, pp. 2797--2810, 2023.

\bibitem{spiralshard}
Y.~Lin, M.~Li, and J.~Zhang, ``Spiralshard: Highly concurrent and secure blockchain sharding via linked cross-shard endorsement,'' \emph{IEEE Transactions on Networking}, pp. 1--16, 2025.

\end{thebibliography}
% \clearpage

% \input{Files/8_appendix}

\vspace{-35pt}
\begin{IEEEbiography}[{\includegraphics[width=1in,height=1.25in,clip,keepaspectratio]{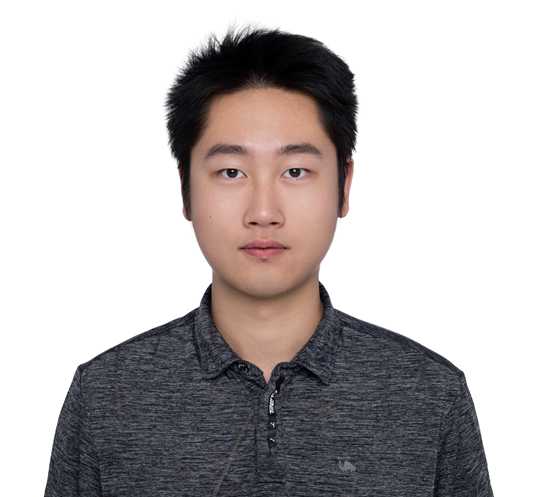}}]{Chaoyue Yin}
is currently a master candidate with Department of Computer Science and Engineering, Southern University of Science and Technology. 
He received his B.E. degree in computer science and technology from Southern University of Science and Technology in 2024. 
His research interests are mainly in blockchain sharding and interoperability protocol.
\end{IEEEbiography}
\vspace{-35pt}
\begin{IEEEbiography}[{\includegraphics[width=1in,height=1.25in,clip,keepaspectratio]{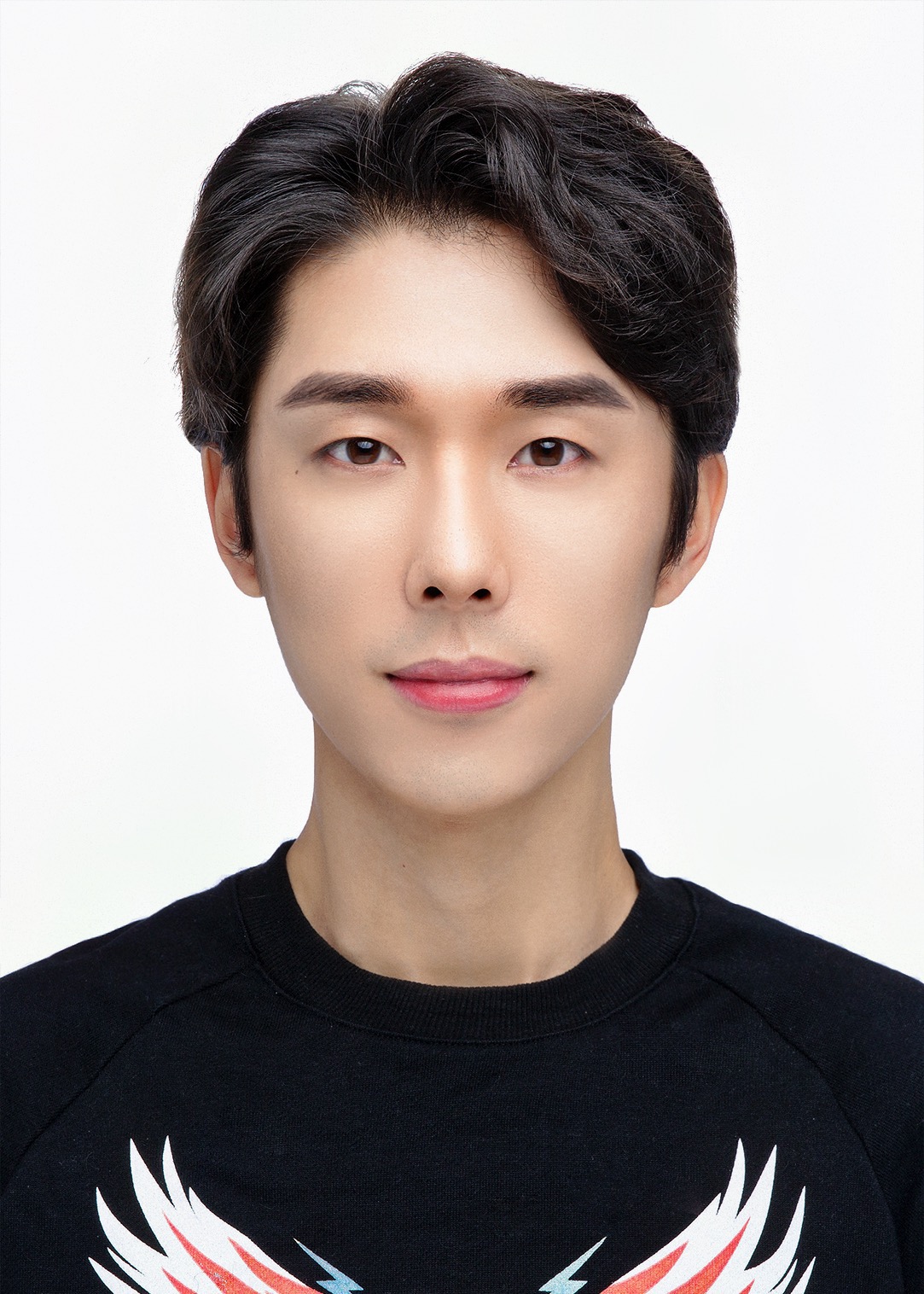}}]{Mingzhe Li}
is currently a US-equivalent Assistant Professor with the School of Computing and Information Technology, Great
Bay University.
% Scientist with the Institute of High Performance Computing (IHPC), A*STAR, Singapore.
He received his Ph.D. degree from the Department of Computer Science and Engineering, Hong Kong University of Science and Technology in 2022.
Prior to that, he received his B.E. degree from Southern University of Science and Technology.
His research interests are mainly in blockchain sharding, consensus protocol, blockchain application, network economics, and crowdsourcing.
\end{IEEEbiography}
\vspace{-35pt}
\begin{IEEEbiography}
[{\includegraphics[width=1in,height=1.25in,clip,keepaspectratio]{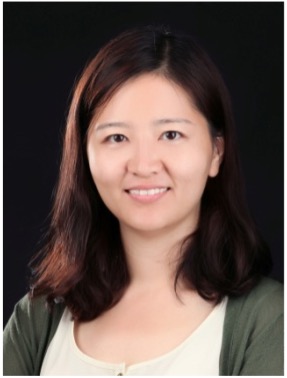}}]{Jin Zhang} 
is currently an associate professor with Department of Computer Science and Engineering, Southern University of Science and Technology. 
She received her B.E. and M.E. degrees in electronic engineering from Tsinghua University in 2004 and 2006, respectively, and received her Ph.D. degree in computer science from Hong Kong University of Science and Technology in 2009. 
% She was then employed in HKUST as a research assistant professor. 
Her research interests are mainly in mobile healthcare and wearable computing, wireless communication and networks, network economics, cognitive radio networks and dynamic spectrum management. 
% She has published more than 50 papers in top-level journals and conferences. 
% She is the Principle Investigator of several research projects funded by National Natural Science Foundation of China, Hong Kong Research Grants Council and Hong Kong Innovation and Technology Commission. 
\end{IEEEbiography}
\vspace{-35pt}
\begin{IEEEbiography}[{\includegraphics[width=1in,height=1.25in,clip,keepaspectratio]{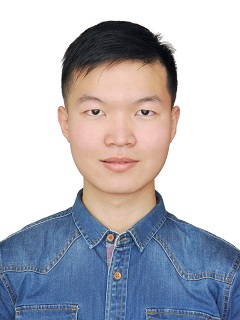}}]{You Lin}
is currently a master candidate with Department of Computer Science and Engineering, Southern University of Science and Technology. 
He received his B.E. degree in computer science and technology from Southern University of Science and Technology in 2021. 
His research interests are mainly in blockchain, network economics, and consensus protocols.
\end{IEEEbiography}
\vspace{-33pt}
\begin{IEEEbiography}[{\includegraphics[width=1in,height=1.25in,clip,keepaspectratio]{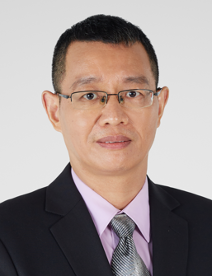}}]{Qingsong Wei}
received the PhD degree in computer science from the University of Electronic Science and Technologies of China, in 2004. He was with Tongji University as an assistant professor from 2004 to 2005. He is a Group Manager and principal scientist at the Institute of High Performance Computing, A*STAR, Singapore. His research interests include decentralized computing, Blockchain and federated learning. He is a senior member of the IEEE.
\end{IEEEbiography}
\vspace{-35pt}
\begin{IEEEbiography}[{\includegraphics[width=1in,height=1.25in,clip,keepaspectratio]{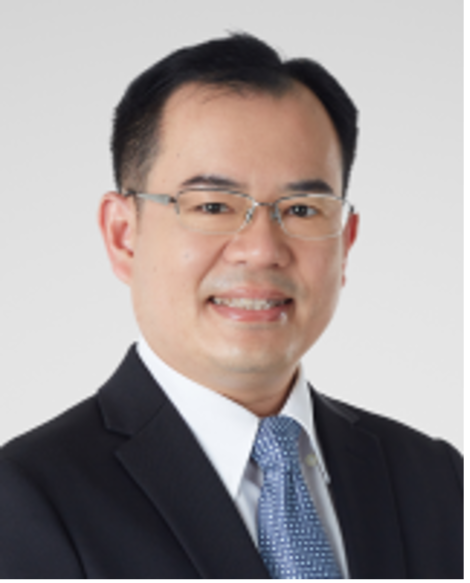}}]{Siow Mong Rick Goh}
received his Ph.D. degree in electrical and computer engineering from the National University of Singapore. He is the Director of the Computing and Intelligence (CI) Department, Institute of High Performance Computing, Agency for Science, Technology and Research, Singapore. His current research interests include artificial intelligence, high-performance computing, blockchain, and federated learning.
\end{IEEEbiography}

\end{document}